\documentclass[aps,showpacs,pre,amsmath,amsfonts,amssymb,superscriptaddress,nofootinbib,notitlepage,a4paper]{revtex4}
\usepackage{amsmath}
\usepackage{amsfonts}
\usepackage{amssymb}
\usepackage{xcolor}
\usepackage{graphicx}
\usepackage{float}
\usepackage{psfrag}
\usepackage{bm}
\usepackage[naturalnames]{hyperref}
\usepackage{enumitem}
\usepackage{algpseudocode}

\newcommand{\ind}[0]{\mathbb{I}}
\newcommand{\bfA}[0]{{\bf A}}
\newcommand{\bfS}[0]{{\bf S}}
\newcommand{\cH}[0]{\mathcal{H}}
\newcommand{\cF}[0]{\widehat{F}}
\newcommand{\cB}[0]{\widehat{B}}

\begin{document}

\bibliographystyle{myunsrt}

\title{Small Coupling Expansion for Multiple Sequence Alignment}

\author{Louise Budzynski}
\affiliation{DISAT, Politecnico di Torino, Corso Duca degli Abruzzi, 24, I-10129, Torino, Italy} 
\affiliation{Italian Institute for Genomic Medicine, IRCCS Candiolo, SP-142, I-10060, Candiolo (TO), Italy}
\author{Andrea Pagnani}
\affiliation{DISAT, Politecnico di Torino, Corso Duca degli Abruzzi, 24, I-10129, Torino, Italy} 
\affiliation{Italian Institute for Genomic Medicine, IRCCS Candiolo, SP-142, I-10060, Candiolo (TO), Italy}
\affiliation{INFN, Sezione di Torino, Torino, Via Pietro Giuria, 1 10125 Torino Italy}

\begin{abstract}
The alignment of biological sequences such as DNA, RNA, and proteins, is one of the basic tools that allow to detect evolutionary patterns, as well as functional/structural characterizations between homologous sequences in different organisms. Typically, state-of-the-art bioinformatics tools are  based on profile models that assume the statistical independence of the different sites of the sequences. Over the last years, it has become increasingly clear that homologous sequences show complex patterns of long-range correlations over the primary sequence as a consequence of the natural evolution process that selects genetic variants under the constraint of preserving the functional/structural determinants of the sequence. Here, we present an alignment algorithm based on message passing techniques that overcomes the limitations of profile models. Our method is based on a perturbative small-coupling expansion of the free energy of the model that assumes a linear chain approximation as the $0^\mathrm{th}$-order of the expansion. We test the potentiality of the algorithm against standard competing strategies on several biological sequences.
\end{abstract}

\maketitle

\section{Introduction} 
The evolution of biological molecules such as proteins is an ongoing highly nontrivial dynamical process spanning over billions of years, constrained by the maintenance of relevant structural, and functional
determinants. One of the most striking features of natural evolution is
how different evolutionary pathways produced ensemble of molecules
characterized by an extremely heterogeneous amino acid sequence --
often with a sequence identity lower than 30\% -- but with virtually
identical three-dimensional native structures. Thanks to a shrewd use
of this structural similarity, it is nowadays possible to classify the
entire set of known protein sequences into disjoint classes of
sequences originating from a common ancestral sequence. Sequences
belonging to the same class are called homologous. 



Homologous sequences are best compared using sequence alignments \cite{durbin1998}. 
Depending on the number of sequences to align, there are three possible options: 
(i) {\em Pairwise Alignments} aims at casting two sequences into the same 
framework. The available algorithms are
typically based on some versions of dynamic programming, and scale linearly 
with the length of the sequences \cite{needleman1970,smith1981}. 
(ii) {\em Multiple Sequence Alignments} (MSA) maximize the global similarity 
of more than two sequences \cite{edgar2006}. 
Dynamic programming techniques can be generalized to more than two
sequences, but with a computational cost that scales exponentially with
the number of sequences to be aligned. Producing MSAs
of more than $10^3$ sequences remains an open computational challenge. 
(iii) To align larger number of homologous sequences, one first selects a representative subset called {\em seed} for which the use of MSA is computationally feasible. Every single homolog eventually is aligned to the {\em seed} MSA. In this way one can easily align up to $10^6$ sequences \cite{altschul1997,eddy2011,el2019}.

Standard alignment methods are based on the {\em independent site evolution} assumption 
\cite{durbin1998}, {\em i.e.} the probability of observing a sequence is factorized among the different
sites. From a statistical mechanics perspective, such approximation
corresponds to a non-interacting 21 colors  (20 amino acids + 1 gap symbol) Potts model. Profile hidden Markov models \cite{eddy2011}, for instance, are of that type. The computational complexity of profile models is polynomial. However, profile models neglect long-range correlations, although they are an important statistical feature of homologous proteins. This well-known phenomenon is at the basis of what biologists call {\em epistasis} ({\em i.e.} how genetic variation depend on the genetic context of the sequence). Recently, epistasis has received renewed attention from the statistical mechanics' community \cite{dejuan2013}. Given an MSA of a specific protein family, one could ask what is the best statistical description of such an ensemble of sequences. Summary statistics such as one-site frequency count $f_i(a)$ (i.e. the empirically observed frequency of observing amino acid $a$ at position $i$ in the MSA), two-site frequency count $f_{ij}(a,b)$ (i.e. the frequency of observing the amino acid realization $a,b$ at position $i$ and $j$ respectively), and in principle higher-order correlations, could be used to inverse statistical modeling of the whole MSA. One can assume that each sequence in the MSA is independently drawn from a multivariate distribution $P(a_1, \dots,a_L)$ constrained to reproduce the multibody empirical frequency counts of the MSA. The use of the maximum-entropy principle is equivalent to assume a Boltzmann-Gibbs probability measure for $P$. The related Hamiltonian is a 21-colors generalized Potts model characterized by two sets of parameters: local fields $H_i(a)$, and epistatic two-site interaction terms $J_{i,j}(a,b)$. Such parameters can be learned more or less efficiently, using the so-called Direct Coupling Analysis (DCA) \cite{cocco2018}. This method has found many interesting applications ranging from the prediction of protein structures \cite{marks2011,morkos2011}, protein-protein interaction \cite{procaccini2011,baldassi2014,feinauer2016}, prediction of mutational effects \cite{figliuzzi2015,morcos2016,hopf2017,trinquier2021}, etc. Inherent to this strategy, there is the counter intuitive step of constructing an MSA based on a statistical independence of sites assumption, which is used, in turn, to predict long-range correlations. To solve this loophole, we propose a mean field message-passing strategy to align sequences to a reference Potts model. To do so, we considered a first-order perturbative expansion {\em a la} Plefka \cite{Plefka82}, setting as $0^{\mathrm{th}}$ order of the expansion the linear chain approximation. Recently, other strategies have been proposed which take into account long range correlations: search for remote homology \cite{eddy2020}, a simplified version of the message-passing strategy presented here \cite{Muntoni2020}, alignment of two Potts models \cite{talibart2021}, and a more machine learning inspired method based on tranformers \cite{ovchinnikov2021}. 

\section{Set-up of the problem}  
\begin{figure}[ht]
	\centering
	\includegraphics[width=0.5\columnwidth]{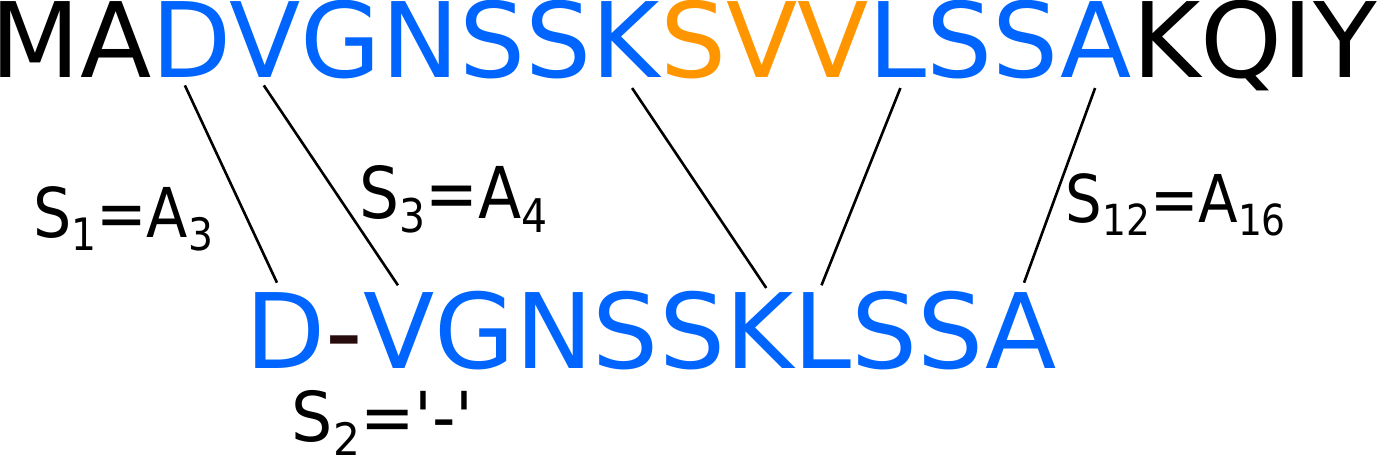}
	\caption{{\bf Example of alignment.} 
		Top: original sequence $\bfA$ of length $N=20$, bottom: aligned sequence $\bfS$ of length $L=12$. Match states are enlightened in (dark gray) blue. There is one gap at position $2$ in the sub-sequence $\bfS$. Three amino acids are skipped in the original sequence (in (light gray) orange): they are interpreted as insertions.
	}
	\label{fig:ex-align}
\end{figure}

Although here we will focus on proteins, the method can be extended to other biological sequences, such as RNA and DNA. Let $\bfA = (A_1, \dots, A_N)$ be an unaligned amino acid sequence of length $N$, containing a protein domain $\bfS = (S_1, \dots, S_L)$ of a known protein family. While $\bfA$ contains only amino acids (represented as upper-case letters from the amino acid alphabet), $\bfS$ might also contain gaps that are used to indicate the deletion of an amino acid in the sequence $\bfA$. We assume that the protein family is described by a Potts Hamiltonian:
\begin{align}
    \cH_{\rm DCA}(\bfS) = -\sum_{i=1}^L H_i(S_i) - \sum_{i<j} J_{ij}(S_i,S_j) \ .
    \label{eq:hamiltonian0}
\end{align}
The couplings $J_{ij}$ and external fields $H_i$ have been learned from the seed MSA in a preprocessing step, using DCA, and the sub-sequence $\bfS$ is assumed to have the same length $L$ as the seed.
The energy $\cH_{\rm DCA}$ is considered as a score for the sub-sequence $\bfS$ to belong to the protein family. 
In this setting, our problem consists in finding a sub-sequence $\bfS$ with the lowest energy (i.e. with the highest score).
Contrarily to profile models, the Hamiltonian $\cH_{\rm DCA}$ also includes pairwise interactions related to residue co-evolution, hopefully leading to more accurate alignments in cases where conservation of single residues is not sufficient to describe the protein family.
The Hamiltonian in Eq.~(\ref{eq:hamiltonian0}) does not model the insertions statistics, because the parameters $J_{ij}$ and $H_i$ are learned from the seed MSA, in which all columns containing inserts have been removed.
Therefore, as  in~\cite{Muntoni2020}, we added the insertion cost $\cH_{\rm ins}$ , which has been learned from the insertion statistics contained in the full seed alignment. 
Similarly to~\cite{Muntoni2020}, we also added an additional gap cost $\cH_{\rm gap}$ to correct the gap statistics learned in $\cH_{\rm DCA}$ (that deeply depends on how the seed is constructed).
In this setting, the alignment problem corresponds to finding a sub-sequence $\bfS=(S_1, \dots, S_L)$ of the original sequence $\bfA=(A_1, \dots, A_N)$, such that:
\begin{enumerate}
	\item $\bfS$ is an ordered list of amino acids in $\bfA$ (called {\em match} states), with the possibility of adding gaps states denoted ``-" between two consecutive positions, and of skipping some amino acids of $\bfA$ (i.e. interpreting them as insertions).
	\item the sub-sequence $\bfS$ minimizes the total energy $\cH = \cH_{\rm DCA} + \cH_{\rm ins} + \cH_{\rm gap}$.
\end{enumerate}
An example of a sequence $\bfA$ and its alignment $\bfS$ is illustrated in Fig.~(\ref{fig:ex-align}).
In order to formulate this problem as a statistical physics model, we introduce for each position $i=1, \dots L$ a pair of variables $y_i=(x_i,n_i)$, where $x_i\in\{0,1\}$ is a binary variable, and $n_i\in\{0, 1, \dots, N, N+1\}$ is a pointer. 
The variable $x_i$ indicates whether position $i$ is a gap $``-"$ ($x_i=0$) or a match state ($x_i=1$).
When $i$ is a match, the pointer $n_i$ indicates the position of the match state in the full-length sequence $\bfA$. When $i$ is a gap, the pointer keeps track of the last match state before position $i$.
Note that we added pointer values $n=0$ and $n=N+1$. These value are used for gap states at the beginning and at the end of the aligned sequence: if matched symbols start to appear only from a position $i>1$, we fill the previous positions $j<i$ with gaps having pointer $n_j=0$. Similarly, if the last matched state appears at position $i<L$, we fill the next positions $j>i$ with gaps having pointers $n_j=N+1$.
The Potts Hamiltonian re-written in terms of the variables ${\bf y} =(y_1, \dots, y_L)$ is:
\begin{align*}
    \cH_{\rm DCA}({\bf y}) = -\sum_{i=1}^L H_i(A_{x_i.n_i}) - \sum_{i<j} J_{ij}(A_{x_i.n_i},A_{x_j.n_j}) \ , 
\end{align*}
where $A_0=-$ is the gap state.
We will use short-hand notations $H_i(y_i)\equiv H_i(A_{x_i.n_i})$ and $J_{ij}(y_i,y_j)\equiv J_{ij}(A_{x_i.n_i},A_{x_j.n_j})$ in the rest of the paper.
The insertion cost $\cH_{\rm ins}$ and the gap cost $\cH_{\rm gap}$ take the form introduced in\cite{Muntoni2020}. In particular for the insertion cost we have:
\begin{align*}
\cH_{\rm ins}({\bf y}) = \sum_{i=2}^L\varphi_i(n_i-n_{i-1}-1) \ , 
\end{align*}
with  $\varphi_i(\Delta n) = (1-\delta_{\Delta n, 0})[\lambda_o^i + \lambda_e^i(\Delta n - 1)]$, and $\Delta n_i=n_i-n_{i-1}-1$ the number of skipped amino acids between position $i-1$ and $i$. The parameters $\{\lambda_o^i, \lambda_e^i\}$ have been inferred from the insertion statistics (see~\cite{Muntoni2020} section IV.B.).
And for the gap cost we have:
\begin{align*}
\cH_{\rm gap}({\bf y}) = \sum_{i=1}^L\mu(x_i, n_i) \ , 
\end{align*}
with $\mu(1,n)=0$ for match states, $\mu(0,0)=\mu(0,N+1)=\mu_{\rm ext}$ for external gaps, and $\mu(0,n)=\mu_{\rm int}$ for internal gaps (with $0<n<N+1$). 
The values of $\mu_{\rm int}$, and $\mu_{\rm ext}$ have been chosen according to the procedure described in~\cite{Muntoni2020}, section IV.C: one re-align sequences of the seed MSA using several values of $\mu_{\rm int}, \mu_{\rm ext}$, and pick the ones minimizing the Hamming distance between the re-aligned seed and the original seed.
 
We finally introduce the Boltzmann probability law over the set of possible alignments:
\begin{align}
    \label{eq:Boltzmann}
    P({\bf y}) = \frac{\chi_{\rm in}(y_1)\prod_{i=2}^L\chi_{\rm sr}(y_{i-1}, y_i)\chi_{\rm end}(y_L)}{Z(\beta)}e^{-\beta \cH({\bf y})},
\end{align}
where $\chi_{\rm in}$, $\chi_{\rm sr}$ and $\chi_{\rm end}$ are Boolean functions ensuring that the ordering constraints are satisfied.
The constraint for $\bfS$ to be an ordered list of amino acids is $\bfA$ can indeed be encoded with the function $\chi_{\rm sr}(x_{i-1},n_{i-1}, x_i, n_i)$ between two consecutive positions:
\begin{align*}
\chi_{\rm sr}(0,n_{i-1}, 0, n_i) &= \ind[n_{i-1}=n_i] \\
\chi_{\rm sr}(1,n_{i-1}, 0, n_i) &= \ind[n_{i-1}=n_i \vee n_i=N+1] \\
\chi_{\rm sr}(0,n_{i-1}, 1, n_i) &= \ind[0\leq n_{i-1}<n_i<N+1] \\
\chi_{\rm sr}(1,n_{i-1}, 1, n_i) &= \ind[0<n_{i-1}<n_i<N+1] \ ,
\end{align*}
and with additional constraints imposed in the first and last position:
\begin{align*}
\chi_{\rm in}(x_1,n_1) &= \delta_{x_1,0}\delta_{n_1,0} + \delta_{x_1,1}\ind[0<n_1<N+1] \\
\chi_{\rm end}(x_L,n_L) &= \delta_{x_L,0}\delta_{n_L,N+1} + \delta_{x_L,1}\ind[0<n_L<N+1] \ .
\end{align*}
Configurations ${\bf y}$ violating the ordering constraints have zero-probability. The parameter $\beta$ plays the role of an inverse-temperature: by increasing $\beta$, the distribution concentrates on the allowed configurations achieving the smallest energy, i.e. on the best alignments.

\section{Small Coupling Expansion} 

\begin{figure}[ht]
	\centering
	\includegraphics[width=0.7\columnwidth]{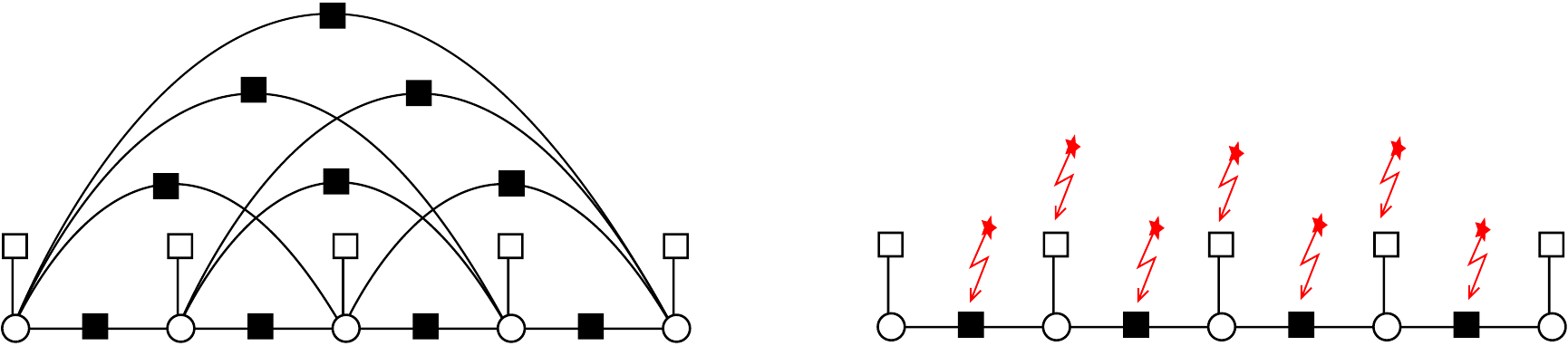}
	\caption{Left panel: Fully-connected factor graph associated to the probability Eq.~(\ref{eq:Boltzmann}) with $L=5$. Variables $y_i$ are represented by white dots, external fields $H_i$ by white squares, and couplings $J_{ij}$ by black squares. 
	Right panel: Factor graph obtained after the perturbative expansion. External fields $H_2, \dots, H_{L-1}$ and short-range couplings $J_{i,i+1}$, $i\in\{1,\dots,L-1\}$ are modified according to Eq.~(\ref{eq:perturbed_couplings_fields}) (illustrated by red (light gray) stars).
	{\bf }
	}
	\label{fig:factorgraphs}
\end{figure}

An efficient strategy for approaching this constrained optimization problem is to use Belief-Propagation (BP).
BP is a message-passing method to approximate probability distributions of the form of Eq.~(\ref{eq:Boltzmann}).
In particular it allows to compute marginal probabilities on any small subset of variables, as well as the partition function $Z(\beta)$. 
BP is exact when the factor graph representing interactions between variables is a tree, and is used as an heuristic for sparse graphs. 
In our case however, the set of couplings $J_{ij}$ is defined for all pairs $(i,j)$, resulting in a fully-connected factor graph, as shown in the left panel of Fig.~\ref{fig:factorgraphs}.
This makes the problem difficult for BP. However, although the interactions are very dense (all couplings are non-zero), they are typically weak for distant sites. Conversely, interactions between two neighbor sites are typically stronger as they encode the one-dimensional structure of the amino acid sequence.

Therefore, in this work we develop an approximation method where long-range couplings are treated perturbatively.
More precisely, we perform a small-coupling expansion of the free-energy $F=-\frac{1}{\beta}\log Z(\beta)$ associated with the Boltzmann distribution in Eq.~(\ref{eq:Boltzmann}), where the zero-th order corresponds to the model defined on the one-dimensional chain, i.e. with long-range couplings set to zero: $J_{ij}=0$ for $|i-j|>1$.
Higher orders take into account the contribution of long-range couplings in a perturbative way. 
We performed the expansion up to the first-order term, and let the computation of higher orders for future work.
This pertubative expansion is similar to a Plefka expansion to obtain the TAP equations \cite{Plefka82, YeGe91,OpMa01}. The main difference is that in the Plefka expansion, the $0^\mathrm{th}$ order is the mean field model (i.e. including only external fields $H_i$) and all couplings $J_{ij}$ are treated perturbatively, while in our approach the $0^\mathrm{th}$ order includes also the short-range couplings $J_{i,i+1}$.
We then study the stationary points of the perturbed free-energy with respect to single-sites and nearest-neighbors sites marginal probabilities $P_i(y_i)$ and $P_{i,i+1}(y_i, y_{i+1})$, to obtain a set of approximate BP equations.
The technical details of this small-coupling expansion are given in appendices~\ref{sec:small_coupling_expansion}. and~\ref{sec:stationarity}.
In the rest of the paper we refer to these approximate BP equations as the Small Coupling Expansion (SCE) equations.

\begin{figure}[ht]
	\centering
	\includegraphics[width=0.7\columnwidth]{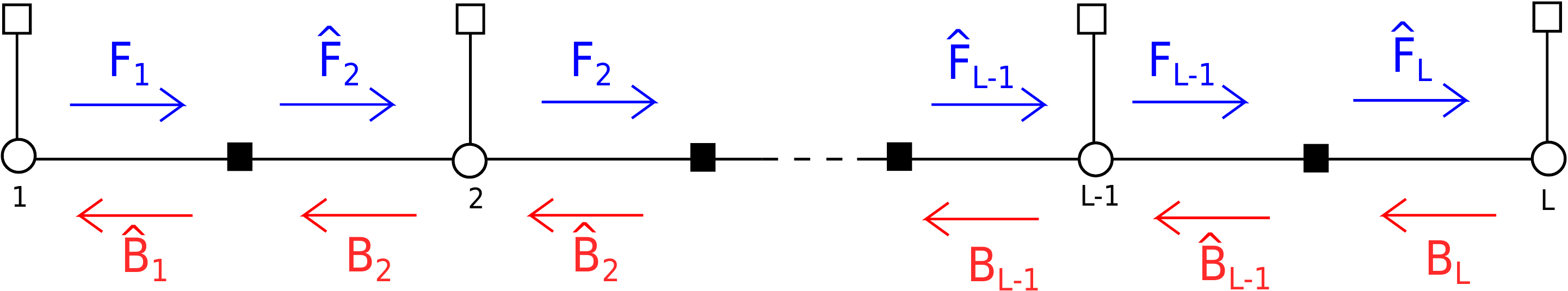}
	\caption{
	{\bf BP messages defined on the one-dimensional chain.}  
	In blue (top arrows): the set of forward messages $F_i,\cF_i$, and in red (bottom arrows) the set of backward messages $B_i,\cB_i$.
	}
	\label{fig:bp-messages}
\end{figure}

This set of SCE equations can be seen as BP equations whose associated factor graph is a linear chain, as represented in the right panel of Fig.~\ref{fig:factorgraphs}, or equivalently to the equations obtained with the transfer matrix method (or dynamic programming/forward-backward algorithm \cite{durbin1998}). The contribution of the long-range couplings $J_{ij}$, $|i-j|>1$ results into a modification of the external fields $H_i$ and short-range couplings $J_{i,i+1}$:
\begin{align}
\label{eq:perturbed_couplings_fields}
    \begin{aligned}
        \widetilde{H}_i &= H_i + f_i \quad \text{for} \quad i\in\{2,\dots,L-1\} \\
        \widetilde{J}_{i,i+1} &= J_{i,i+1} + g_{i} \quad \text{for} \quad i\in\{1, \dots, L-1\} \ .
    \end{aligned}
\end{align}
Single-site fields $f_i$ and nearest-neighbors pairwise fields $g_{i}$ are computed explicitly from the set of conditional probabilities $P(y_i|y_j)$ for any $i,j$ with $|i-j|>1$:
\begin{align}
\label{eq:expression_f_main}
    f_l(y_l) = -\sum_{i=1}^{l-1}\sum_{j=l+1}^L\sum_{y_i,y_j}J_{ij}(y_i,y_j)P_i(y_i|y_l)P_j(y_j|y_l)
\end{align}
and:
\begin{align}
\label{eq:expression_g_main}
    g_{l}(y_l, y_{l+1}) &= \sum_{i=1}^{l}\sum_{j=\zeta_i^l}^L \sum_{y_i,y_j}J_{ij}(y_i,y_j)  P_i(y_i|y_l)P_j(y_j|y_{l+1}) 
\end{align}
with $\zeta_i^l=\max(l+1,i+2)$. 
The SCE equations are recursive equations for a set of {\it forward} messages $F_i(y_i),\cF_i(y_i)$, and {\it backward} messages $B_i(y_i),\cB_i(y_i)$, defined on the edges of the one-dimensional chain, as shown in Fig.~\ref{fig:bp-messages}.
We give here the exact form of the approximate BP equations, their derivation is given in appendix~\ref{sec:stationarity}. For the forward messages we have:
\begin{align}
\begin{aligned}
\label{eq:BP_update_forward}
F_1(y_1) &= \frac{1}{z_{1\to e_1}}e^{\beta H_1(y_1)}  \\
F_i(y_i) &= \frac{1}{z_{i\to e_i}}e^{\beta \widetilde{H}_i(y_i) }\cF_i(y_i) \ , \ \text{for} \quad i\geq 2  \\
\cF_{i+1}(y_i) &=\frac{1}{\widehat{z}_{e_i\to i+1}} \sum_{y_i} e^{ \beta \widetilde{J}_{e_i}(y_i, y_{i+1}) } F_i(y_i) \ ,
\end{aligned}
\end{align}
where $F_i$ is defined for $i\in\{1, \dots, L-1\}$ and $\cF_i$ for $i\in\{2,\dots, L-1\}$, and $z_{i\to e_i}$, $\widehat{z}_{e_i\to i+1}$ are normalization factors ensuring that the BP messages are normalized to $1$. And for the backward messages we have:
\begin{align}
\begin{aligned}
\label{eq:BP_update_backward}
B_L(y_L) &= \frac{1}{z_{L\to e_{L-1}}}e^{\beta H_L(y_L)}  \\
B_i(y_i)&=\frac{1}{z_{i\to e_{i-1}}}e^{\beta \widetilde{H}_i(y_i) }\cB_i(y_i) \ , \ \text{for} \quad i\leq L \\
\cB_i(y_i) &=\frac{1}{\widehat{z}_{e_i\to i}} \sum_{y_{i+1}} e^{ \beta \widetilde{J}_{e_i}(y_i, y_{i+1})  } B_{i+1}(y_{i+1}) \ ,
\end{aligned}
\end{align}
where $B_i$ is defined for $i\in\{2, \dots, L\}$ and $\cB_i$ for $i\in\{1,\dots, L-2\}$, and $z_{i\to e_{i-1}}, \widehat{z}_{e_i\to i}$ are normalization constants.
Single-site and nearest-neighbors marginal probabilities $P_i(y_i)$ and $P_{i,i+1}(y_i, y_{i+1})$ can be expressed in terms of the BP messages:
\begin{align}
\label{eq:BP_marginals_single}
\begin{aligned}
P_1(y_1) &=\frac{1}{z_1}e^{\beta H_1(y_1)} \cB_1(y_1) \\
P_i(y_i) &=\frac{1}{z_i} e^{\beta\widetilde{H}_i(y_i)} \cF_i(y_i)\cB_i(y_i) \ , \ 2\leq i\leq L-1 \\
P_L(y_L) &=\frac{1}{z_L}e^{\beta H_L(y_L)} \cF_L(y_L) \ ,
\end{aligned}
\end{align}
and for $i\in\{1,\dots, L-1\}$:
\begin{align}
\label{eq:BP_marginals_nn}
\begin{aligned}
P_{i,i+1}(y_i, y_{i+1}) &= \frac{e^{\beta \widetilde{J}_{i,i+1}(y_i, y_{i+1})}}{z_{i,i+1}} F_i(y_i) B_{i+1}(y_{i+1})\ .
\end{aligned}
\end{align}

From the set of marginal probabilities, one finally computes the conditional probabilities $P_i(y_i|y_j)$, for all $i,j$ with $|i-j|>1$, from the chain rule, which is valid when long-range couplings are neglected:
\begin{align}
\label{eq:conditional_recursion_main}
    P_i(y_i|y_l) = \sum_{y_{i-1}}P_{i-1}(y_{i-1}|y_l)P_i(y_i|y_{i-1}) \quad  \text{if} \ i>l+1 \ ,
\end{align}
with a similar expression similarly when $i<l-1$.

A solution the SCE equations can be found iteratively (see appendix~\ref{subsec:SCE_algo}. for a complete description of the algorithm). From a random initialization of the BP messages, the algorithm first computes the marginals $P_i$, $P_{i,i+1}$ from Eq.~(\ref{eq:BP_marginals_single}, \ref{eq:BP_marginals_nn}), then updates the set of conditional probabilities $P_i(y_i|y_j)$ from Eq.~(\ref{eq:conditional_recursion_main}), and finally computes the long-range fields $f_i, g_{i}$ using Eqs.~(\ref{eq:expression_f_main}, \ref{eq:expression_g_main}). BP messages are then updated using the new value of $f_i, g_{i}$, and these steps are repeated until convergence.
Each iteration has complexity $O(L^3Q^4)$, with $Q$ the size of state space for variable $y_i$ (in our case $Q=2(N+2)$), the bottleneck being the computation of fields $f_i, g_{i}$. 
Although this algorithm is slower than DCAlign, the approximate BP algorithm derived in~\cite{Muntoni2020}, it has the advantage to derive the small coupling expansion in a rigorous way, which in turns allows to compute thermodynamic quantities such as free-energy and entropy (see appendix~\ref{sec:thermo_quantities}. for their explicit expression), that were not available with the previous approach~\cite{Muntoni2020}.
The free-energy could be used to optimize the Hamiltonian's parameters (in particular the gap costs $\mu_{\rm int}, \mu_{\rm ext}$ defined in $\cH_{\rm gap}$).
We leave this for future work.
Note that DCAlign equations~\cite{Muntoni2020} can be recovered from this perturbative expansion, at the cost of assuming the factorization $P_{ij}(y_i,y_j)\simeq P_i(y_i)P_j(y_j)$ for $|i-j|>1$ in the first-order term of the free-energy (see appendix~\ref{subsec:recover_DCAlign}. for an explicit derivation).

\subsection{Decoding strategies.}
Once a solution to the SCE equations is found, an assignment can be computed from the marginals using a decoding strategy.
We used and compared the performance of two strategies: (i) {\it nucleation} already used in~\cite{Muntoni2020}, (ii) and {\it Viterbi decoding} in which we use the nearest-neighbors pairwise marginals $P_{i,i+1}$ to compute the solution having the largest probability of being generated by a Markov chain using transition probabilities $P(y_{i+1}|y_i)$. 
Note that neither of the two strategies are guaranteed to produce an assignment achieving the largest probability w.r.t. Eq.~(\ref{eq:Boltzmann}): in principle one should use a decimation strategy and re-compute after each assignment of a variable the new marginals conditioned on the previous assignments. 
However, these two strategies are faster than decimation, and we have seen that they provide very good alignments.
In particular, we show below that Viterbi decoding outperforms the nucleation strategy on protein families PF00684 and PF00035 taken from the Pfam database (https://www.ebi.ac.uk/interpro/ release 32.0 release 32.0)~\cite{el2019}.
More details on the decoding strategies are given in appendix~\ref{sec:decodings}.

\section{Epsilon Coupling Analysis}
\begin{figure*}
	\centering
	\includegraphics[width=1.0 \columnwidth]{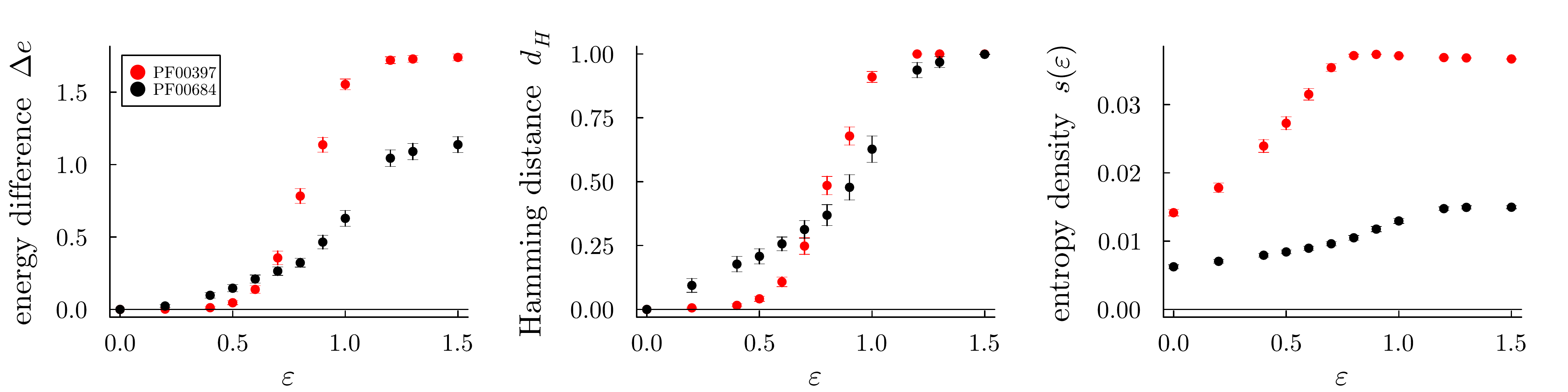}
	\caption{{\bf Result for $\epsilon$-coupling.} Red (light gray) points: on family PF00397, averaged over $100$ sequences.
		Black points: on family PF00684, averaged over $40$ sequences.
		Left: Difference between energy densities of the ground state: $\Delta e = (\cH({\bf y}^{\epsilon})-\cH({\bf y}^0))/L$. 
		Middle: Hamming distance between the ground state at $\epsilon$ and at $\epsilon=0$. 
		Right: Entropy density $s(\epsilon). 
		$
		Results are obtained with SCE + Viterbi decoding, with an annealed scheme $\beta\in\{0, 0.05, \dots, 0.4\}$.
	}
	\label{fig:epsilon_coupling}
\end{figure*}
The SCE approach allows us to find a solution to the constrained optimization problem of finding the best alignment of the original sequence $\bfA$ to a seed MSA. We used this method to explore the energy landscape around a given optimal alignment found with our algorithm.
We used a general technique called the Epsilon Coupling Analysis, introduced in\cite{PagParRat03}, see also \cite{MaPaRi02} for its application to RNA secondary structures: starting from the optimal solution ${\bf y}^0$, we add a repulsive external field to the Hamiltonian $\cH({\bf y})$, that repel ${\bf y}^0$ with intensity $\epsilon$:
\begin{align}
\label{eq:epsilon_hamiltonian}
	\cH'({\bf y};\epsilon, {\bf y}^0) = \cH({\bf y}) + \epsilon\sum_{i=1}^L\delta_{y_i , y^0_i}\ ,
\end{align}
This additional term -- {\em viz.} the Hamming distance $d_H({\bf y}, {\bf y}^0)$ between the optimal solution and a configuration ${\bf y}$ --
penalizes structures that are close to the ground state ${\bf y}^0$, allowing to explore other minima. One computes the optimal solution ${\bf y}^{\epsilon}$ of $\cH'$, for many values of $\epsilon$, using again the SCE + decoding strategy.
For each value of $\epsilon$, one compares the new ground state with the true one by computing their Hamming distance $d_H({\bf y}^0, {\bf y}^{\epsilon})$, and their difference in energy density $\Delta e = (\cH({\bf y}^{\epsilon})-\cH({\bf y}^0))/L$.
We also compute for each value of $\epsilon$, the entropy density $s(\epsilon)$ associated with the perturbed model~(\ref{eq:epsilon_hamiltonian}).
Results are shown in Fig.~\ref{fig:epsilon_coupling} for two protein families (PF00397 and PF00684) selected from the Pfam database~\cite{el2019}.
We restricted our analysis to short families ($L=67$ for PF00684 and $L=31$ for PF00397) in order to avoid a significant slowing down of the alignment algorithm.
As $\epsilon$ increases, ${\bf y}^{\epsilon}$ starts to depart from ${\bf y}^0$ ($d_H>0$) and simultaneously the difference in energy density $\Delta e$ becomes positive. This indicates that we do not find other optimal solutions, instead we find solutions with higher energy ($\Delta e >0$), but close in hamming distance to the true ground state, suggesting a landscape with a single minimum in a basin of attraction. This analysis is compatible with our computation of the entropy: we obtain for both families a rough estimate of the number of optimal configurations $e^{Ls(\epsilon)}$ between $1$ and $2$ configurations. At  larger $\epsilon$ values, the energy density difference $\Delta e$, the Hamming distance $d_H$ and the entropy $s(\epsilon)$ reach a plateau at $\epsilon\simeq 1.0$ for both protein families. The solutions ${\bf y}^{\epsilon}$ found for these values of $\epsilon$ are mostly made of gaps, i.e. are not good alignments, which indicates that in this regime the free energy landscape has been substantially modified by the perturbation.

\section{Performance analysis}
\begin{figure*}
	\centering
	\includegraphics[width=1.0\columnwidth]{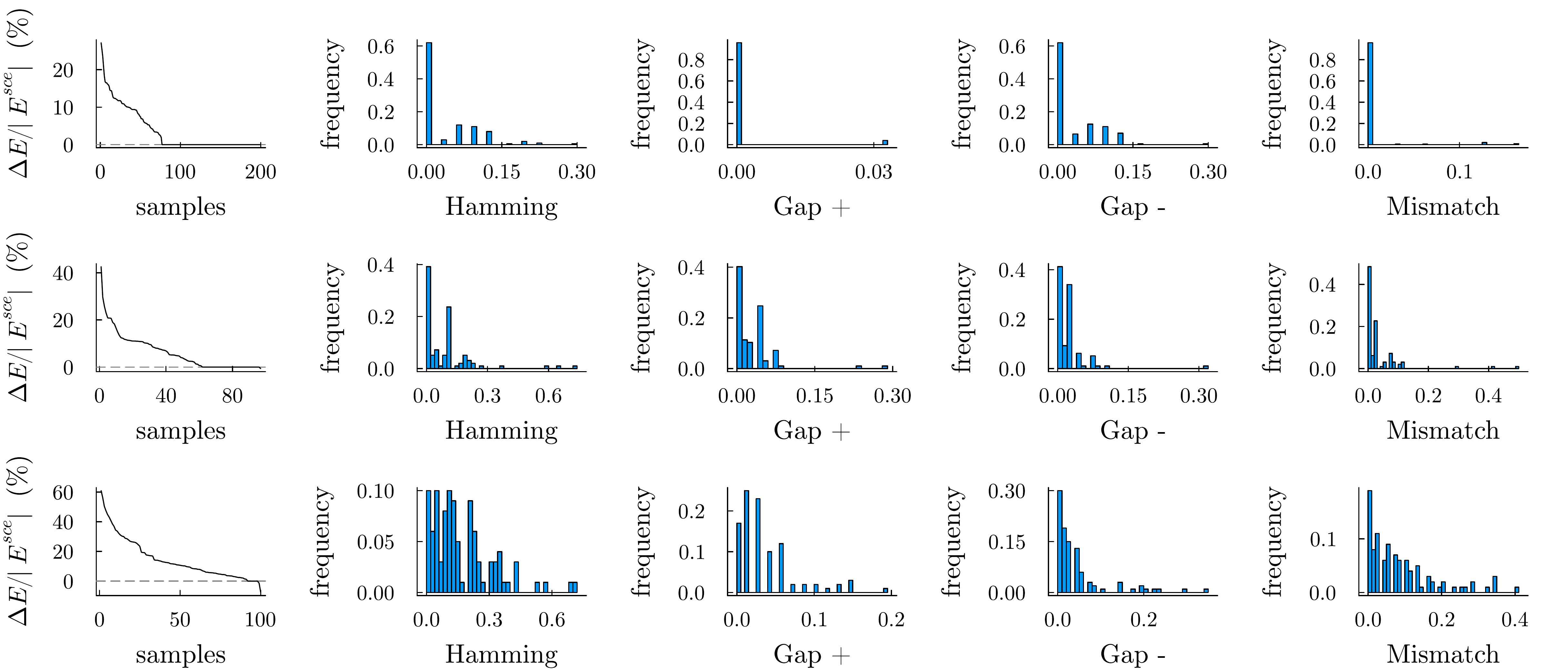}
	\caption{{\bf Comparison of SCE with HMMER.} Top: protein family PF00397 (on a set of $200$ sequences). Middle: protein family PF00684 (on a set of $100$ sequences). Bottom: protein family PF00035 (on a set of $100$ sequences). On left panels, we plot difference in energy between the ground state found with HMMER $E^{\rm hmmer}=\cH_{\rm DCA}(\bfS^{\rm hmmer})$ and the ground state found with SCE $E^{\rm sce}=\cH_{\rm DCA}(\bfS^{\rm sce})$ (percent of the ground state energy $E^{\rm sce}$ found with SCE). Positive $\Delta E = E^{\rm hmmer} - E^{\rm sce}$ means that SCE strategy has found a better (lower in energy) alignment than HMMER. Samples are sorted by decreasing values of $\Delta E/|E^{\rm sce}|$.
		Then, from left to right, we plot the histograms of Hamming distances, Gap +, Gap - and Mismatch.
	}
	\label{fig:seqs_hmmerSCE}
\end{figure*}
We assessed the quality of MSAs generated by our SCE method, and compared them to state-of-the art alignments provided by HMMER\cite{eddy2011}, on small protein families PF00397, PF00684 and PF00035 taken from Pfam~\cite{el2019} (with $L=67$ for PF00035).
As done in~\cite{Muntoni2020}, we did not consider the entire sequences, whose length $N$ is often much larger than $L$, but a “neighbourhood” of the hit selected by HMMER. In practice we add $\delta$ amino acids at the beginning and at the end of the hit resulting in a final length $N = \delta + L + \delta$ (with $\delta=20$ for PF00397 and PF00684, and $\delta=10$ for PF00035).
We consider sequence-wise measures, also used in~\cite{Muntoni2020}, to evaluate the similarity between two candidate MSAs (a ``reference" and a ``target" MSA): 
(i) {\bf Hamming} distance between two alignments ($\bfS^{\rm ref}$ and $\bfS^{\rm tar}$) of the same sequence $\bfA$ in the reference and target MSAs respectively. (ii) {\bf Gap} $\boldsymbol{+}$: Number of match states in $\bfS^{\rm ref}$, that have been replaced by a gap in $\bfS^{\rm tar}$. (iii) {\bf Gap} ${\boldsymbol -}$: Number of gap states in $\bfS^{\rm ref}$, that have been replaced by a match state in $\bfS^{\rm tar}$. (iv) {\bf Mismatch}: Number of amino acid mismatches, i.e. the number of times we have a match state in both $\bfS^{\rm ref}$ and $\bfS^{\rm tar}$, but corresponding to different amino acids positions in the full sequence $\boldsymbol{A}$. All quantities are normalized by $L$, the length of the sequences.

In addition, we compare the quality of alignments by computing for each sequence of the MSAs the difference in energy density $\Delta e = (\cH_{\rm DCA}(\bfS^{\rm ref}) - \cH_{\rm DCA}(\bfS^{\rm tar}))/L$.

\subsection{Comparison with HMMER}
We first compare the MSA produced by our SCE algorithm (target MSA) with the MSA produced by HMMER (reference MSA), see Fig.~\ref{fig:seqs_hmmerSCE}. For each family we choose a random sample of sequences and compare the alignments produced by the two methods.
The difference in energy density for each sequence (sorted in decreasing order) is plotted on the left panels. 
For the three families, we see that for a large fraction of the sample set, the energy $\cH_{\rm DCA}$ of the SCE alignment is lower than the one of HMMER, thus resulting in a better alignment found by SCE. 
For PF00397 and PF00684, for the rest of the sample set, the difference in energy is zero: both methods have found the same alignment.
The distribution of similarity metrics (Hamming distance, Gap$\pm$, Mismatch) are mostly concentrated on the first bins for both families, indicating that alignments found by SCE and HMMER are close.
For PF00035, it is only on a tiny fraction of the sample set that SCE finds a solution with either equal or slightly higher energy compared to HMMER. The distribution of similarity metrics is broader on this family, indicating that SCE and HMMER find substantially different alignments on a large fraction of samples.

\subsection{Comparison with the seed}
To explore further the differences between our method and HMMER, we compared the alignments found with the two methods and the seed MSA. More precisely we have re-aligned each sequence of the seed MSA (reference MSA) with our method and with HMMER to obtain a new MSA (target MSAs).
Results - given in appendix~\ref{subsec:comp_seed}. Fig.~\ref{fig:seqs_seedVS_SCEandHmmer}. (for the protein family PF00397) - show that the MSA obtained with SCE is closer to the seed MSA than the one obtained with HMMER, suggesting that SCE is performing better the alignment task. 

\subsection{Comparison of decoding methods}
We  compared the performances of two decoding methods: {\it nucleation} and {\it Viterbi} (see appendix~\ref{sec:decodings}.).
For each family, we compare the two decoding methods used on the set of marginal probabilities computed from our SCE algorithm.
Results are shown in appendix~\ref{subsec:comp_decodings}, Fig.~\ref{fig:seqs_decodings} (for families PF00397, PF00684 and PF00035). 
While for family PF00397, both decoding methods find essentially the same alignment, the situation is different for families PF00684 and PF00035: although for a large fraction of the sequences, both decoding methods find the same alignment, we can clearly see that Viterbi finds a better solution on a non-negligible fraction the sequences, with a significantly lower energy, and nucleation leads to a better alignment only for a few sequences. 

\subsection{Remote homology detection}
\begin{figure}
	\centering
	\includegraphics[width=0.5\columnwidth]{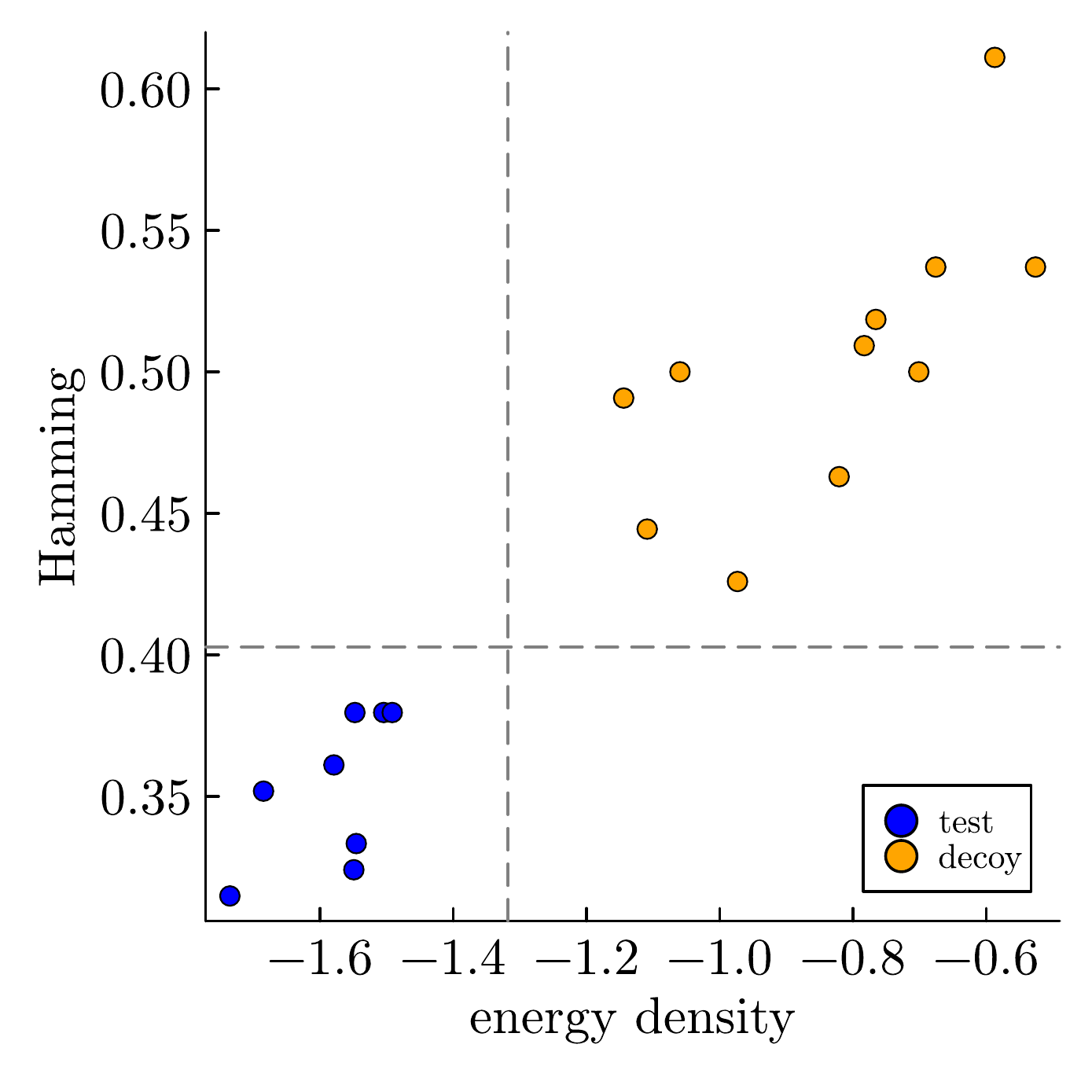}
	\caption{{\bf Remote homology detection on the RNA family RF00162.} \cite{WilburnEddy_remote20}
		Alignments found by SCE+Viterbi decoding, on a set of $8$ test sequences (in blue/dark gray) and $11$ decoy sequences (in orange/light gray). 
		$x$-axis: energy density $e=\cH(\bfS)/L$ of the alignment.
		$y$-axis: Hamming distance from the solution to closest aligned sequence in the training set.
		Vertical (resp. horizontal) dashed line shows the average between the right-most (resp. highest) blue point and the left-most (resp. lowest) orange point, indicating that the two sets can be separated with both observables.
	}
	\label{fig:remote-homology-RF00162}
\end{figure}
We test the performance of our SCE algorithm on homology search for the RNA family RF00162 taken from Rfam database~\cite{Rfam_17}.
The goal of homology detection is to determine whether a sequence is evolutionary related (i.e. homologous) to a family of sequences.  It is common that homology search fails at identifying distantly related sequences~\cite{Eddy_homologyfailure}.
As testing ground, we use the SAM riboswitch seed alignment from Rfam-family~\cite{Rfam_17} RF00162 (which have length $L=108$). This data-set has been proposed in~\cite{eddy2020} as a stress-test for alignment algorithms. 
Following this set-up, the MSA is  divided into a training set, and a test set. Sequences in the test set are selected in order to be distant to the training set and distant one from each other (see~\cite{eddy2020} for details). In addition, a set of non-homologous decoy sequences is randomly generated as follows: each character is drawn i.i.d. from the nucleotide composition of the positive test sequences, with a length matching a randomly selected positive test sequence~\cite{eddy2020}. To wrap up, we have three mutually non overlapping set of sequences: (i) training: (from which we learn the parameters of our model), (ii) test: a set of homologous sequences, (iii) decoy: a randomly generated set of non-aligned sequences.

For each sequence in the test set and for $11$ sequences randomly extracted from the set of decoy sequences, we compute the alignment found with SCE + Viterbi decoding. The parameters are learned from the training set: the parameters of the Potts model are trained with a Boltzmann machine DCA learning algorithm, and the parameters of the insertion cost $\cH_{\rm ins}$ are learned from the insertion statistics (see~\cite{Muntoni2020} section IV.B.). The parameters of the gap cost $\mu_{\rm int}, \mu_{\rm ext}$, are taken from~\cite{Muntoni2020}, Table.II.

To score the alignments, we compare their energy density $e=\cH(\bfS)/L$. We also compute, for each alignment $\bfS$ found by our algorithm, its Hamming distance w.r.t. each aligned sequence in the training set.
We then collect the minimum attained value. Results are given in Fig.~\ref{fig:remote-homology-RF00162}., and show that our method is able to disentangle between decoy sequences and true sequences belonging to RF00162: the alignments found for the test set have smaller energy, and are closer to the training set. 

\section{Conclusion}
We proposed an alternative method based on a perturbative expansion of the model around the linear chain, and obtained a set of approximate message-passing equations that we used to find optimal alignments. 
We tested the potentiality of our algorithm on protein families taken from the Pfam database~\cite{el2019}. 
The results obtained on these families suggest that including long-range correlations is crucial for the alignment task, and it is a promising direction to go beyond current state-of-the-art bioinformatics tools based on profile models which, from a statistical mechanics standpoint, are assuming statistical independence of sites. Additionally, we compare the performances of two different decoding strategies, and show that for two of the protein families studied in this paper, Viterbi decoding algorithm outperforms the nucleation strategy presented in~\cite{Muntoni2020}. 
We test the performance of our method on remote homology search, for the RNA family RF00162 taken from the Rfam database~\cite{Rfam_17}, and obtain promising results suggesting that our method is able to detect distant homologs.
The method proposed in this paper treats perturbatively the contribution of long-range couplings $J_{ij}$, with $|i-j|>1$ using a small coupling expansion {\em a la} Plefka \cite{Plefka82}. While this assumption might not be justified as some of the couplings might not be in the perturbative regime, our approach is the first step to include them, in order to go beyond the independent-site assumption. Moreover, in the context of DCA \cite{morkos2011}, it has been empirically shown that the first order approximation of the Plefka expansion is enough to capture relevant structural and functional features of the protein family.

Our approach provides a self-consistent derivation of the {\em naive} mean-field approximation used in~\cite{Muntoni2020}, which, in turn, being variational, allows us to compute approximated thermodynamic potentials.
We use this new strategy to explore the free-energy landscape of this constrained optimization problem, obtaining, for the protein families studied in this paper, the global picture of a unique solution surrounded by a basin of attraction. The main limitation of our method is an increase of computational complexity with respect to the {\em naive} mean-field method~\cite{Muntoni2020}: indeed, our SCE algorithm has an $O(L^3N^4)$ complexity (with $L$ the length of the alignment $\bfS$ and $N$ the length of sequence $\bfA$ to be aligned), compared to the $O(L^2N^2)$ complexity for the DCAlign algorithm designed in~\cite{Muntoni2020}. Further investigations could be to use our approximation of the free-energy for developing methods to simultaneously optimize the model's parameters and find the optimal alignment, using for instance strategies based on expectation-maximization. Note finally that the method developed in this paper is not restricted to the alignment problem, and could be used in other problems that have the structure of a one-dimensional chain with additional fully-connected weak couplings.

\acknowledgments
The authors thank Anna Paola Muntoni for interesting discussions and for sharing with us part of the code needed for data processing. AP acknowledges funding by the EU H2020 research and innovation programme MSCA-RISE-2016 under Grant Agreement No. 734439 INFERNET, as well as for financial support from FAIR (Future Artificial Intelligence Research) and ICSC (Centro Nazionale di Ricerca in High-Performance Computing, Big Data, and Quantum Computing) founded by European Union – NextGenerationEU.

\appendix
\section{Performance analysis}
\label{sec:performance_analysis}
\subsection{Comparison with the seed}
\label{subsec:comp_seed}
\begin{figure*}[ht]
	\centering
	\includegraphics[width=1.0\columnwidth]{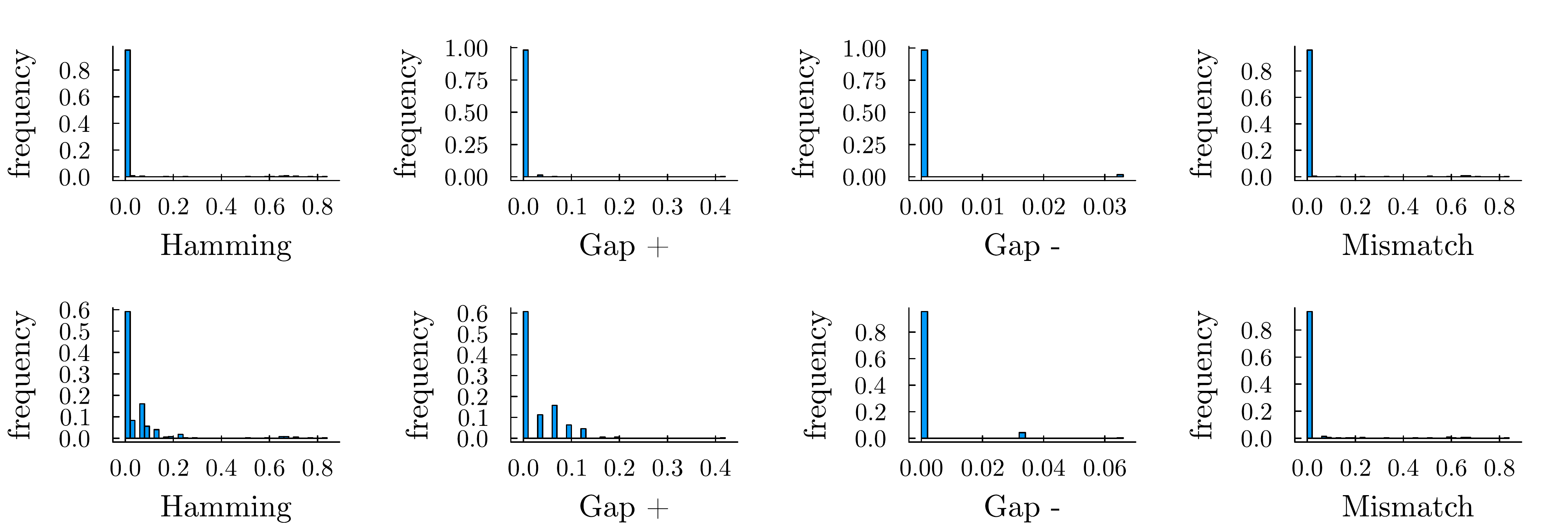}
	\caption{{\bf Comparison between the seed MSA and the MSAs produced with SCE and HMMER}, for PF00397. Top: comparison with SCE. All similarity distributions are concentrated on zero: SCE finds the same alignment as the one of the seed for most of the sequences. Bottom: comparison with HMMER. The distribution has more weight on non-zero distances: HMMER finds alignments more distant to the seed than the alignments found by SCE.
	}
	\label{fig:seqs_seedVS_SCEandHmmer}
\end{figure*}

Fig.~\ref{fig:seqs_seedVS_SCEandHmmer} gives the results obtained when comparing the MSAs produced by a given alignment strategy (SCE+decoding and HMMER) with the seed MSA, on protein family PF00397. 
Each sequence present in the seed MSA is re-aligned with our SCE algorithm (top panels) and with HMMER (bottom panels). 
One evaluates the similarities between the produced MSA and the seed: high similarity means that the alignment algorithm performs well.

\subsection{Comparison of the decoding methods}
\label{subsec:comp_decodings}
\begin{figure*}[ht]
	\centering
	\includegraphics[width=1.0\columnwidth]{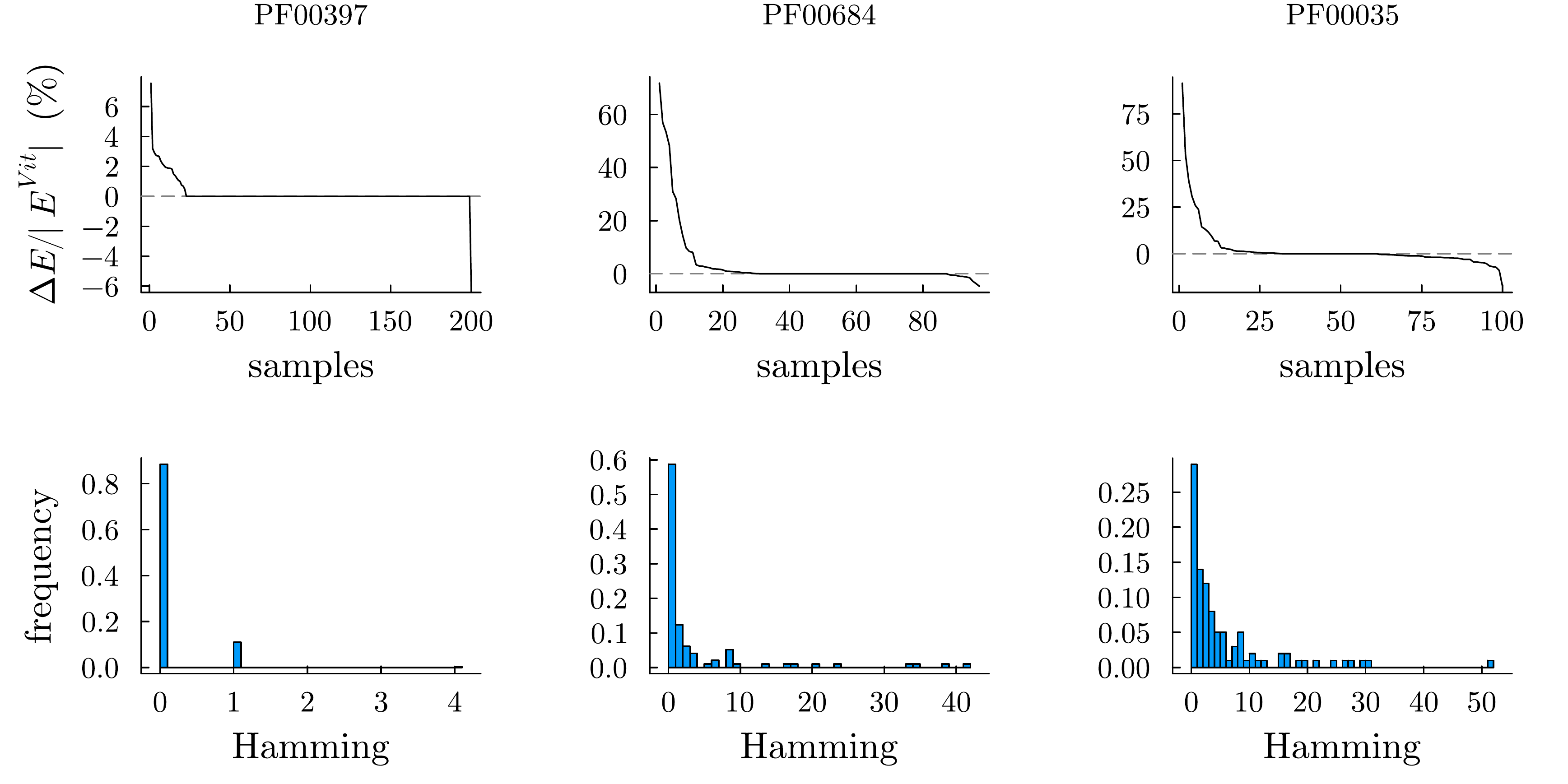}
	\caption{{\bf Comparison of decoding methods:} From left to right, results shown for protein families PF00397, PF00684, PF00035. 
	Top panels: difference in energy between the ground state found with nucleation $E^{\rm nucl}=\cH_{\rm DCA}(\bfS^{\rm nucl})$ and the ground state found with Viterbi $E^{\rm Vit}=\cH_{\rm DCA}(\bfS^{\rm Viterbi})$ (percent of the ground state energy $E^{\rm Vit}$ found with Viterbi). Positive $\Delta E = E^{\rm nucl} - E^{\rm Vit}$ means that Viterbi decoding has found a better (lower in energy) alignment than nucleation. Samples are sorted by decreasing values of $\Delta E/|E^{\rm Vit}|$.
	Bottom panels: Histogram of the hamming distances (number of differing components) between the ground states found with nucleation and with Viterbi.
	}
	\label{fig:seqs_decodings}
\end{figure*}

Fig.~\ref{fig:seqs_decodings} shows the comparison between nucleation and Viterbi decoding methods for protein families PF00397, PF00684 and PF00035.  
On PF00397, Viterbi and nucleation strategies find essentially the same alignments, as one can see from the histogram of Hamming distances between the two alignments: they are either identical, or differ on only one component.
The situation is different on families PF00684 and PF00035, where the histogram of Hamming distance has more weights on larger distances. One can see that Viterbi finds a better alignment on a representative fraction of the samples, with significantly lower energy: up to $\Delta E/|E^{\rm Vit}|=60\%$ for PF00684 and $75\%$ for PF00035 (with $\Delta E = E^{\rm nucl} - E^{\rm Vit}$, and $E^{\rm nucl},E^{\rm Vit}$ the energy of the solutions found with nucleation and Viterbi respectively). 
The marginals needed to perform the decoding are computed with our SCE algorithm. 
Note that we can also use Viterbi decimation on marginals computed with DCAlign algorithm~\cite{Muntoni2020}, we checked that it provides similar results on these families (data not shown).

\section{Small Coupling Expansion}
\label{sec:small_coupling_expansion}
Consider the Hamiltonian:
\begin{align*}
    H({\bf y}) = -\sum_{i=1}^L H_i(y_i) - \sum_{i<j}J_{ij}(y_i, y_j) \ .
\end{align*}
over a set of $L$ variables ${\bf y}=\{y_1,\dots,y_L\}$, with $y_i\in\chi$ a given state space. Note that the Potts model for sequence alignment studied in this paper can be written in this form by including the hard constraints $\chi_{\rm sr}, \chi_{\rm in}, \chi_{\rm end}$, and the insertion $\cH_{\rm ins}$ and gap costs $\cH_{\rm gap}$ inside the external fields $H_i$ and couplings $J_{i, i+1}$.
We want to treat perturbatively the long-range couplings $J_{ij}$, with $|i-j|>1$.
We thus re-write the above Hamiltonian, separating long-range and short-range couplings, and introducing a small parameter $\alpha$:
\begin{align}
    \label{eq:hamiltonian_alpha}
    H({\bf y}) = -\sum_{i=1}^L H_i(y_i) - \sum_{i=1}^{L-1}J_{i,i+1}(y_i, y_{i+1}) - \alpha \sum_{i=1}^{L-2}\sum_{j={i+2}}^L J_{ij}(y_i, y_j) \ .
\end{align}
The approach that we adopted is to perform a perturbative expansion to the first order in $\alpha$ of the free energy associated with this Hamiltonian, in a way similar to the Plefka's expansion \cite{Plefka82, YeGe91}, see also \cite{OpMa01}.

Following the steps of \cite{OpMa01} (chapter 2, section 7), we define a variational free-energy associated to $H({\bf y})$:
\begin{align}
    \label{eq:variational_F}
    \begin{aligned}
    F(Q) &= E(Q) -\frac{1}{\beta}S(Q) \ , \ \text{with} \\
    E(Q) &= \sum_{{\bf y}}Q({\bf y})H({\bf y}) \ , \\ 
    S(Q) &= -\sum_{{\bf y}}Q({\bf y})\log Q({\bf y}) \ .
    \end{aligned}
\end{align}
We know that the minimum of $F(Q)$ is achieved for $Q=P_\beta$, with $P_\beta$ the Boltzmann distribution: $$P_\beta({\bf y})=\frac{e^{-\beta H({\bf y})}}{Z(\beta,\alpha)} \ .$$
The approach consists in doing this minimization in two steps. 
In the first step we perform a constrained minimization in the family of all distributions, which match the single site marginals, and the marginals on all pairs of neighbors sites: 
\begin{align}
    \label{eq:marginal_constraints}
    \begin{aligned}
    Q_i(y_i)&=P_{\beta,i}(y_i) \quad \text{for} \ i\in\{1,\dots,L\} \ ,  \ y_i\in \chi   \\
    Q_{i,i+1}(y_i, y_{i+1})&=P_{\beta, i,i+1}(y_i, y_{i+1}) \quad \text{for}  \ i\in\{1,\dots,L-1\} \ , \ y_i,y_{i+1}\in \chi \ .
    \end{aligned}
\end{align}
We define the Gibbs free energy as the constrained minimum:
\begin{align}
    \label{eq:Gibbs_potential}
    \begin{aligned}
        G(\{P_i\}_{i\in[1,L]}, \{P_{i,i+1}\}_{i\in[1,L-1]}) = \min_{Q}\left\{F(Q): Q_i=P_i \  \forall i\in\{1,\dots,L\}, \ Q_{i,i+1}=P_{i,i+1}\ \forall i\in\{1,\dots,L-1\}\right\} \ .
    \end{aligned}
\end{align}
In the second step, we minimize $G$ over the set of functions $\underline{P}=\{\{P_i\}_{i\in[1,L]}, \{P_{i,i+1}\}_{i\in[1,L-1]}\}$. 
Since the minimizer of $F(Q)$ is the Boltzmann distribution $P_\beta$, we know that the minimizer of $G$ is the set of true marginals: \{$P_{\beta,i}\}_{i\in[1,L]}, \{P_{\beta,i,i+1}\}_{i\in[1,L-1]}$.
To perform the constrained optimization (\ref{eq:Gibbs_potential}) we introduce a set of Lagrange multipliers $\underline{\lambda}=\{\{\lambda_i\}_{i\in[1,L]}, \{\lambda_{i,i+1}\}_{i\in[1,L-1]}\}$.
We then need to minimize the functional:
\begin{align}
    \mathcal{L}(Q) &= F(Q) - \sum_{i=1}^L\sum_{y_i}\lambda_i(y_i)(Q_i(y_i)-P_i(y_i)) - \sum_{i=1}^{L-1}\sum_{y_i, y_{i+1}}\lambda_{i,i+1}(y_i,y_{i+1})(Q_{i,i+1}(y_i, y_{i+1})-P_{i,i+1}(y_i, y_{i+1})) \ ,
\end{align}
where the Lagrange multipliers $\{\lambda_i\}_{i\in[1,L]}, \{\lambda_{i,i+1}\}_{i\in[1,L-1]}$ must be chosen in such a way that the set of constraints (\ref{eq:marginal_constraints}) is satisfied.
The minimizing distribution is:
\begin{align}
    Q_{\underline{\lambda}}({\bf y}) = \frac{1}{Z(\beta, \alpha, \underline{\lambda})}e^{-\beta H({\bf y}) + \beta\sum_{i=1}^L\lambda_i(y_i) + \beta\sum_{i=1}^{L-1}\lambda_{i,i+1}(y_i, y_{i+1})} \ , 
\end{align}
with $Z(\beta, \alpha, \underline{\lambda})$ a normalization constant.
Plugging this solution into the expression (\ref{eq:Gibbs_potential}) of $G$ we get:
\begin{align}
    G(\underline{p}, \underline{\lambda})=\sum_{i=1}^L\sum_{y}\lambda_i(y)P_i(y) + \sum_{i=1}^{L-1}\sum_{y,y'}\lambda_{i,i+1}(y,y')P_{i,i+1}(y,y') - \frac{1}{\beta}\log Z(\beta, \alpha, \underline{\lambda}) \ .
\end{align}
The condition on the Lagrange multipliers is finally obtained by looking at the stationary points of $G$ with respect to the $\lambda_i(y)$'s, $\lambda_{i,i+1}(y,y')$'s. 
Let $\underline{\lambda}^*(\alpha)$ be a set of Lagrange multipliers achieving the stationary point, we then have $$G(\underline{p})=G(\underline{p}, \underline{\lambda}^*(\alpha))\ ,$$ where we have emphasized the dependence in $\alpha$ of the Lagrange multipliers.

We can now perform a perturbation expansion of $G(\underline{p})$ in the small coupling parameter $\alpha$. 
At the first order this gives:
\begin{align}
    G(\alpha) = G(0) + \alpha\left .\frac{{\rm d} G}{{\rm d} \alpha}\right |_{\alpha=0} \quad .
\end{align}
The zeroth order can be computed easily:
\begin{align}
\begin{aligned}
    G(0)&=\sum_{i=1}^L\sum_{y}\lambda^*_i(y)P_i(y) + \sum_{i=1}^{L-1}\sum_{y,y'}\lambda^*_{i,i+1}(y,y')P_{i,i+1}(y,y') -\frac{1}{\beta}\log Z(\beta, \underline{\lambda}^*(\alpha=0))   \\
    &= \sum_{{\bf y}}Q_{\underline{\lambda}^*(\alpha=0)}({\bf y})H({\bf y}; \alpha=0) - \frac{1}{\beta}S(Q_{\underline{\lambda}^*(\alpha=0)}) \\
    &= -\sum_{i=1}^L\sum_y H_i(y)P_i(y) - \sum_{i=1}^{L-1}\sum_{y,y'}J_{i,i+1}(y,y')P_{i,i+1}(y,y') - \frac{1}{\beta}S(Q_{\underline{\lambda}^*(\alpha=0)}) \ ,
\end{aligned}
\end{align}
where in the second line we have used the usual identity between entropy and free-energy for the distribution $Q_{\underline{\lambda}^*(\alpha=0)}$, and in the third line we replaced the Hamiltonian $H({\bf y})$ at $\alpha=0$ by its expression (\ref{eq:hamiltonian_alpha}).
Note that at $\alpha=0$, the interactions occurring in the distribution $Q_{\underline{\lambda}^*(\alpha=0)}$ are lying on the one-dimensional chain, therefore its entropy can be expressed exactly in terms of the marginals $\{\{P_i\}_{i\in[1,L]}, \{P_{i,i+1}\}_{i\in[1,L-1]}\}$:
\begin{align}
    S(Q_{\underline{\lambda}^*(\alpha=0)}) = -\sum_{i=1}^L(1-d_i)\sum_y P_i(y)\log P_i(y) - \sum_{i=1}^{L-1}\sum_{y,y'}P_{i,i+1}(y,y')\log P_{i,i+1}(y,y')
\end{align}
with $d_i$ the degree of site $i$ on the chain (i.e. $d_1=d_L=1$, and $d_i=2$ for $i\in\{2,\dots,L-1\})$. 
The computation of the first order term can also be done, the terms containing the derivative with respect to the $\lambda_i, \lambda_{i,i+1}$ cancel out, and we get:
\begin{align}
\label{eq:first_order_G}
    \left.\frac{{\rm d} G}{{\rm d} \alpha}\right|_{\alpha=0} = -\sum_{i=1}^{L-2}\sum_{j=i+2}^L\sum_{y_i, y_j}J_{i,j}(y_i,y_j)Q_{\underline{\lambda}^*(\alpha=0),i,j}(y_i,y_j) \ .
\end{align}
Finally, we note that at $\alpha=0$, the joint distribution $Q_{\underline{\lambda}^*(\alpha=0),i,j}(y_i,y_j)$ can be expressed in terms of the single site marginals $P_i$'s and short-range marginals $P_{i,i+1}$'s using the chain rule:
\begin{align}
    Q_{\underline{\lambda}^*(\alpha=0),i,j}(y_i,y_j) = \sum_{y_{i+1},\dots,y_{j-1}}\frac{\prod_{k=i}^{j-1}P_{k,k+1}(y_k,y_{k+1})}{\prod_{k=i+1}^{j-1}P_k(y_k)} \ .
\end{align}

We can therefore express the first-order expansion of the free energy only in terms of the single site marginals $P_i$'s and short-range marginals $P_{i,i+1}$ as 
\begin{align}
\label{eq:free_energy_1storder}
    F = F^{\rm sr} - \alpha \mathcal{A} \ , 
\end{align}
with $F^{\rm sr}$ the Bethe free-energy of the model at $\alpha=0$:
\begin{align}
	\label{eq:shortrange-freeen}
\begin{aligned}
    -\beta F^{\rm sr} &= -\sum_{i=1}^L(1-d_i)\sum_y P_i(y)\log P_i(y) - \sum_{i=1}^{L-1}\sum_{y,y'}P_{i,i+1}(y,y')\log P_{i,i+1}(y,y')  \\
    & + \beta \sum_{i=1}^L \sum_y P_i(y) H_i(y) + \beta \sum_{i=1}^{L-1} \sum_{y,y'} P_{i,i+1}(y,y') J_{i,i+1}(y,y') \ ,
\end{aligned}
\end{align}
and with the first-order correction:
\begin{align}
    \mathcal{A} &= \sum_{i=1}^{L-2}\sum_{j=i+2}^L\sum_{y_i, y_j}J_{i,j}(y_i,y_j)\sum_{y_{i+1},\dots,y_{j-1}}\frac{\prod_{k=i}^{j-1}P_{k,k+1}(y_k,y_{k+1})}{\prod_{k=i+1}^{j-1}P_k(y_k)} \ .
\end{align}

In the next section we look at the stationarity of this expression with respect to the marginals $P_i, P_{i,i+1}$'s, and extract from them a set of message-passing equations.
For now on we also set $\alpha=1$, and consider that the long-range couplings are themselves small compared to the next-neighbors couplings: $J_{ij} \ll J_{i',i'+1}$ and $J_{ij} \ll H_{i'}$ for all $i', i, j$ with $|i-j|>1$.

\section{Stationarity of the Free-Energy and Message-Passing equations}
\label{sec:stationarity}
\subsection{Derivation of the equations}
We define the following functions:
\begin{align}
\label{eq:def_external_fields}
\begin{aligned}
    f_i(y) &= \frac{\delta \mathcal{A}}{\delta P_i(y)}  \quad {\rm for} \ i\in\{2,\dots,L-1\} \\
    g_{i}(y,y') &= \frac{\delta \mathcal{A}}{\delta P_{i,i+1}(y,y')}  \quad {\rm for} \ i\in\{1,\dots,L-1\} \ .
\end{aligned}
\end{align}
Note that the functional $\mathcal{A}$ does not depends on the marginals $P_1$ and $P_L$. We search for the stationary points of the functional (\ref{eq:free_energy_1storder}) under the following set of constraints ensuring the local consistency of the marginals (normalization and marginalization):
\begin{align}
\label{eq:marginal_normalization}
\begin{aligned}
    \sum_y P_i(y)&=1 \\
    \sum_y P_{i,i+1}(y,y')&= P_{i+1}(y') \\
    \sum_{y'}P_{i,i+1}(y,y')&=P_i(y) \quad .
\end{aligned}
\end{align}
At this point, one can directly see that the derivative $\frac{\delta F}{\delta P_i(y_i)}$ contains the term $-H_i(y_i)-f_i(y_i)$, while the derivative $\frac{\delta F}{\delta P_{i,i+1}(y_i,y_{i+1})}$ contains the term $-J_{i,i+1}(y_i,y_{i+1})-g_{i}(y_i,y_{i+1})$.
We therefore introduce the functions:
\begin{align}
\begin{aligned}
	\widetilde{H}_i(y_i) &= H_i(y_i)+f_i(y_i) \\
	\widetilde{J}_{i,i+1}(y_i,y_{i+1}) &= J_{i,i+1}(y_i,y_{i+1})+g_{i}(y_i,y_{i+1}) \ ,
\end{aligned}
\end{align}
such that looking at the stationary points of the functional $F$ is equivalent to look at the stationary points of the short-range free-energy $F^{\rm sr}$ (\ref{eq:shortrange-freeen}) with external fields $H_1, \widetilde{H}_2,\dots \widetilde{H}_{L-1},H_L$ and short-range couplings $\widetilde{J}_{1,2},\dots,\widetilde{J}_{L-1,L}$. 

The stationary point of $F^{\rm sr}$ leads to the BP equations on the chain, its derivation can be found for instance in\cite{MeMo09} (where it is done for a generic factor graph). For completeness we recall the main steps here.
We use the shorthand notation $e_i=(i,i+1)$ for the edge linking the two neighbor sites $i,i+1$. Following the steps of \cite{MeMo09} (section 14.4.1.) we introduce Lagrange multipliers $\{\zeta_i\}_{i=1,\dots,L}$ and $\{\zeta_{e_i}^{(i)},\zeta_{e_i}^{(i+1)}\}_{i=1,\dots, L-1}$ ensuring the constraints (\ref{eq:marginal_normalization}) and define the following Lagrangian:
\begin{align}
\begin{aligned}
    \mathcal{L}(\underline{p}, \underline{\zeta}) &= -\beta F(\underline{p}) - \sum_{i=1}^L\zeta_i\left[\sum_y P_i(y)-1\right] - \sum_{i=1}^{L-1}\sum_y\zeta_{e_i}^{(i)}(y)\left[\sum_{y'}P_{e_i}(y,y') - P_i(y)\right] \\
    &- \sum_{i=1}^{L-1}\sum_{y'}\zeta_{e_i}^{(i+1)}(y')\left[\sum_y P_{e_i}(y,y') - P_{i+1}(y')\right] \ ,
\end{aligned}
\end{align}
The stationary point of $\mathcal{L}$ with respect to $P_i$ for $i\in\{2,\dots,L-1\}$ leads to:
\begin{align}
\label{eq:stationarity_beliefs_single}
    P_i(y) = \frac{1}{z_i}\exp\left(-\beta \widetilde{H}_i(y) - \zeta_{e_i}^{(i)}(y) - \zeta_{e_{i-1}}^{(i)}(y) \right) \ ,
\end{align}
where $z_i$ is a constant ensuring the normalization. For $i=1$ and $i=L$ we get the following equations:
\begin{align}
\begin{aligned}
    \beta H_1(y) - \zeta_1 + \zeta_{e_1}^{(1)}(y)&=0 \\
    \beta H_L(y) - \zeta_L + \zeta_{e_{L-1}}^{(L)}(y)&=0 \ .
\end{aligned}
\end{align}
In addition, the stationarity with respect to $P_{e_i}$ gives for $i\in\{1,\dots,L-1\}$:
\begin{align}
    \label{eq:stationarity_beliefs_pair}
\begin{aligned}
    P_{e_i}(y,y') =\frac{1}{z_{e_i}} \exp\left(\beta \widetilde{J}_{e_i}(y,y') - \zeta_{e_i}^{(i)}(y) - \zeta_{e_i}^{(i+1)}(y') \right) \ .
\end{aligned}
\end{align}
The Lagrange multipliers must be chosen in such a way that the constraints (\ref{eq:marginal_normalization}) are satisfied.

We now define the variable-to-factor messages for $i\in\{1, \dots, L-1\}$:
\begin{align}
\begin{aligned}
    \nu_{i\to e_i}(y_i) &=\frac{1}{z_{i\to e_i}} \exp(-\zeta_{e_i}^{(i)}(y_i)) \\
    \nu_{i+1 \to e_i}(y_{i+1}) &=\frac{1}{z_{i+1\to e_i}} \exp(-\zeta_{e_i}^{(i+1)}(y_{i+1}) ) \ ,    
\end{aligned}
\end{align}
and the factor-to-variable messages for $i\in\{1,\dots,L-1\}$:
\begin{align}
\begin{aligned}
    \widehat{\nu}_{e_i\to i}(y_i) &=\frac{1}{\widehat{z}_{e_i\to i}} \sum_{y_{i+1}} \exp( \beta \widetilde{J}_{e_i}(y_i, y_{i+1})  - \zeta_{e_i}^{(i+1)}(y_{i+1}) ) \\
    \widehat{\nu}_{e_i\to i+1}(y_{i+1}) &=\frac{1}{\widehat{z}_{e_i\to i+1}} \sum_{y_i} \exp( \beta \widetilde{J}_{e_i}(y_i, y_{i+1}) - \zeta_{e_i}^{(i)}(y_i) )
\end{aligned}
\end{align}
For compactness we adopt the lighter notations also used in the main text: $F_i=\nu_{i\to e_i}$, $\cF_i=\widehat{\nu}_{e_{i-1}\to i}$ for the forward messages, and $B_i = \nu_{i\to e_{i-1}}$, $\cB_i = \widehat{\nu}_{e_i\to i}$ for the backward messages.
We can directly check that such messages satisfy the following relations:
\begin{align}
\begin{aligned}
    \cB_i(y_i) &=\frac{1}{\widehat{z}_{e_i\to i}} \sum_{y_{i+1}} e^{ \beta \widetilde{J}_{e_i}(y_i, y_{i+1})  } B_{i+1}(y_{i+1}) \\
    \cF_{i+1}(y_i) &=\frac{1}{\widehat{z}_{e_i\to i+1}} \sum_{y_i} e^{ \beta \widetilde{J}_{e_i}(y_i, y_{i+1}) } F_i(y_i)
\end{aligned}
\end{align}
And also:
\begin{align}
\begin{aligned}
    F_1(y_1) &= \frac{1}{z_{1\to e_1}}e^{\beta H_1(y_1)} \\
    B_L(y_L) &= \frac{1}{z_{L\to e_{L-1}}}e^{\beta H_L(y_L)}
\end{aligned}
\end{align}
Further, for each $i\in\{2,\dots,L-1\}$ we have:
\begin{align}
\begin{aligned}
    \sum_{y_{i-1}}P_{e_{i-1}}(y_{i-1},y_i) &\propto B_i(y_i)\cF_i(y_i) \\
    &\propto e^{-\beta \widetilde{H}_i(y_i) }B_i(y_i)F_i(y_i)
\end{aligned}
\end{align}
where in the first line we have used the expression (\ref{eq:stationarity_beliefs_pair}) of $P_{e_{i-1}}$ and the definition of the messages, and in the second line we have used the marginalization condition (\ref{eq:marginal_normalization}) along with the expression of $P_i$ obtained in (\ref{eq:stationarity_beliefs_single}).
We then obtain by eliminating $B_i(y_i)$ from the above relation that for each $i\in\{2,\dots, L-1\}$:
\begin{align}
    F_i(y_i) = \frac{1}{z_{i\to e_i}}e^{\beta \widetilde{H}_i(y_i) }\cF_i(y_i) \ .
\end{align}
A similar relation can be obtained for $B_i$, with $i\in\{2,\dots, L-1\}$:
\begin{align}
    B_i(y_i)=\frac{1}{z_{i\to e_{i-1}}}e^{\beta \widetilde{H}_i(y_i) }\cB_i(y_i) \ ,
\end{align}
We have finally obtained a set of message-passing equations:
\begin{align}
\begin{aligned}
    \label{eq:BP_update}
    F_1(y_1) &= \frac{1}{z_{1\to e_1}}e^{\beta H_1(y_1)}  \\
    F_i(y_i) &= \frac{1}{z_{i\to e_i}}e^{\beta \widetilde{H}_i(y_i) }\cF_i(y_i) \ , \ \text{for} \quad i\in\{2,\dots, L-1\}  \\
    \cF_{i+1}(y_i) &=\frac{1}{\widehat{z}_{e_i\to i+1}} \sum_{y_i} e^{ \beta \widetilde{J}_{e_i}(y_i, y_{i+1}) } F_i(y_i) \ , \ \text{for} \quad i\in\{2,\dots,L\} \\
    B_L(y_L) &= \frac{1}{z_{L\to e_{L-1}}}e^{\beta H_L(y_L)}  \\
    B_i(y_i)&=\frac{1}{z_{i\to e_{i-1}}}e^{\beta \widetilde{H}_i(y_i) }\cB_i(y_i) \ , \ \text{for} \quad i\in\{2, \dots, L-1\} \\
    \cB_i(y_i) &=\frac{1}{\widehat{z}_{e_i\to i}} \sum_{y_{i+1}} e^{ \beta \widetilde{J}_{e_i}(y_i, y_{i+1})  } B_{i+1}(y_{i+1}) \ , \ \text{for} \quad i\in\{1, \dots, L-1\}
\end{aligned}
\end{align}
The single-site marginals $P_i(y_i)$ and the nearest-neighbors pairwise marginals $P_{i,i+1}(y_i, y_{i+1})$ can be expressed in terms of the BP messages: 
\begin{align}
\label{eq:BP_marginals}
\begin{aligned}
    P_1(y_1) &=\frac{1}{z_1}e^{\beta H_1(y_1)} \cB_1(y_1) \\
    P_i(y_i) &=\frac{1}{z_i} e^{\beta\widetilde{H}_i(y_i)} \cF_i(y_i)\cB_i(y_i) \ , \ \text{for} \quad i\in\{2, \dots, L-1\} \\
    P_L(y_L) &=\frac{1}{z_L}e^{\beta H_L(y_L)} \cF_L(y_L) \\
    P_{i,i+1}(y_i, y_{i+1}) &= \frac{1}{z_{i,i+1}}e^{\beta \widetilde{J}_{i,i+1}(y_i, y_{i+1})} F_i(y_i) B_{i+1}(y_{i+1}) \ , \ \text{for} \quad i\in\{1, \dots, L-1\} \ .
\end{aligned}
\end{align}

\subsection{Explicit expression of the long-range fields}
The long-range fields $f_i$ and $g_{i}$ admit an explicit expression in terms of the marginals $P_i, P_{i,i+1}$.
We get for $f_l$, $l\in\{2,\dots, L-1\}$:
\begin{align*}
    f_l(y)&=\frac{\delta \mathcal{A}(\underline{p})}{\delta P_l(y)} \\
    &= -\sum_{i=1}^{l-1}\sum_{j=l+1}^L\sum_{y_i,y_j}J_{ij}(y_i,y_j)\sum_{y_{i+1},\dots,y_{j-1}}\frac{\prod_{k=i}^{l-1}P_{k,k+1}(y_k,y_{k+1})}{\prod_{k=i+1}^{l}P_k(y_k)}\times\frac{\prod_{k=l}^{j-1}P_{k,k+1}(y_k,y_{k+1})}{\prod_{k=l}^{j-1}P_k(y_k)}
\end{align*}
We can express the above expression in terms of conditional probabilities, and obtain:
\begin{align}
\label{eq:expression_f}
    f_l(y_l) = -\sum_{i=1}^{l-1}\sum_{j=l+1}^L\sum_{y_i,y_j}J_{ij}(y_i,y_j)P_i(y_i|y_l)P_j(y_j|y_l)
\end{align}
Similarly, we have for $g_{l}$, $l \in\{1, \dots, L-1\}$:
\begin{align*}
    &g_{l}(y_l,y_{l+1}) = \frac{\delta \mathcal{A}(\underline{p})}{\delta P_{l,l+1}(y_l,y_{l+1})} \\
    &=\sum_{i=1}^l\sum_{j=l+1}^L \mathbb{I}[j>i+1]\sum_{y_i,y_j}J_{ij}(y_i,y_j)\sum_{y_{i+1},\dots y_{j-1}}\frac{\prod_{k=i}^{l-1}P_{k,k+1}(y_k,y_{k+1})}{\prod_{k=i+1}^{l}P_k(y_k)}\times\frac{\prod_{k=l+1}^{j-1}P_{k,k+1}(y_k,y_{k+1})}{\prod_{k=l+1}^{j-1}P_k(y_k)}
\end{align*}
with the convention that $\prod_{k=i}^j = 1$ if $i>j$. We obtain an expression in terms of conditional probabilities:
\begin{align}
\label{eq:expression_g}
    g_{l}(y_l, y_{l+1}) &= \sum_{i=1}^l\sum_{j=l+1}^L \mathbb{I}[j>i+1]\sum_{y_i,y_j}J_{ij}(y_i,y_j) P_i(y_i|y_l)P_j(y_j|y_{l+1})
\end{align}
In the implementation of the algorithm we used the fact that conditional probabilities can be computed recursively on the chain, in order to reduce the number of operation needed:
\begin{align}
\label{eq:conditional_recursion}
\begin{aligned}
    P_i(y_i|y_l) = \sum_{y_{i-1}}P_{i-1}(y_{i-1}|y_l)P_i(y_i|y_{i-1}) \ , \quad \text{if} \quad i>l+1 \\
    P_i(y_i|y_l) = \sum_{y_{i+1}}P_{i+1}(y_{i+1}|y_l)P_i(y_i|y_{i+1}) \ , \quad \text{if} \quad i<l-1
\end{aligned}
\end{align}

\subsection{SCE Algorithm}
\label{subsec:SCE_algo} 
A solution to the set of SCE equations can be obtained with an iterative algorithm whose main steps are described below:
\begin{algorithmic}[1]
	\State Initialize randomly the BP messages.
	\State Compute the marginals $P_1,\dots, P_L$ and $P_{1,2}, \dots, P_{L-1,L}$ from their expression in terms of BP messages eq.~(\ref{eq:BP_marginals}).
	\State Compute the set of conditional probabilities $P_i(y_i|y_j)$, $i,j$ with $|i-j|>1$ from equations (\ref{eq:conditional_recursion}).
	\State Compute the long-range external fields $f_2,\dots, f_{L-1}$ and $g_{1},\dots,g_{L-1}$ using equation (\ref{eq:expression_f}) and (\ref{eq:expression_g}) respectively.
	\State Compute the new BP messages from eq.~(\ref{eq:BP_update}).
	\State Repeat steps 2. to 5. until convergence.
\end{algorithmic}
To enhance the convergence of the above algorithm we used a damping when computing the new BP messages, keeping a fraction $\gamma$ of the old BP messages. In practice we used $\gamma=0.2$ for most of the sequences, and increased its value up to $\gamma=0.9$ for unconverged ones. 

\subsection{Recovering the DCAlign small-coupling approximation}
\label{subsec:recover_DCAlign} 
It is interesting to note that the small-coupling approximation done in \cite{Muntoni2020} for the DCAlign algorithm, can be recovered from the first-order expansion of the free-energy obtained in the previous section.
We recall the expression (\ref{eq:first_order_G}) of the first-order term in the perturbation expansion:
\begin{align*}
    \left.\frac{{\rm d} G}{{\rm d} \alpha}\right|_{\alpha=0} = -\sum_{i=1}^{L-2}\sum_{j=i+2}^L\sum_{y_i, y_j}J_{i,j}(y_i,y_j)P(y_i,y_j)
\end{align*}
The approximate message-passing equations obtained in\cite{Muntoni2020} can be recovered by assuming that the pairwise marginals $P(y_i,y_j)$ for far away sites (i.e. $|i-j|>1$) can be approximated by $$P(y_i,y_j)\approx P(y_i)P(y_j) \ .$$
Under this assumption the external fields (\ref{eq:def_external_fields}) become:
\begin{align}
\begin{aligned}
    & f_i(y_i)= \sum_{j\notin\{i-1,i,i+1\}}\sum_{y_j} J_{ij}(y_i,y_j)P(y_j) \ , \quad  \ \text{for} \quad i\in\{1, \dots, L\}\\
    & g_{i}(y_i,y_{i+1})=0 \ , \quad  \ \text{for} \quad i\in\{1, \dots, L-1\}
\end{aligned}    
\end{align}
Plugging this form into the BP equations one recovers the equations obtained in \cite{Muntoni2020} (section III.B).

\section{Decoding}
\label{sec:decodings}
Once a fixed-point of the message-passing equations (\ref{eq:BP_update}) is found, our aim is to extract a configuration ${\bf y}^*=\{y_1^*,\dots,y_L^*\}$ sampled from the Boltzmann distribution $P_{\beta}({\bf y})$.
When $\beta$ is large enough, the distribution concentrates on configuration achieving minimal energy, which corresponds to optimal alignments.
We used three different methods to decode.

In the first one, we simply take the maximum of each singles-site marginal: 
$$y_i^*=\underset{y}{\rm argmax}\{P_i(y)\} \quad \text{for each} \quad i\in\{1,\dots,L\} \ .$$
This approach is not guaranteed to provide a configuration ${\bf y}^*$ that satisfy the ordering constraints ensured by the functions $\chi_{\rm sr}, \chi_{\rm in}, \chi_{\rm end}$. To overcome this problem, we have used more elaborated decimation procedures described below. Note however that when $\beta$ is large, one expects the single-site marginal to be concentrated on a single-value, and therefore, for most instances, we actually obtain a configuration satisfying all the constraints with this method.

The second method is the nucleation method, already used in \cite{Muntoni2020} (section III.C.), that we recall here.
One first selects the most polarized site 
$$i^*=\underset{i}{\rm argmax}\{\max_{y}(P_i(y)\}$$
and set $y_{i^*}=\underset{y}{\rm argmax}\{P_{i^*}(y)\}$.
Then one extracts the variables $y_{i^*\pm 1}$ as the ones achieving the maximum of the marginals, with the constraint that $y_{i^*\pm 1}$ should be consistent with the choice of $y_{i^*}$:
$$
    y_{i^*+1} = \underset{y}{\rm argmax}\{P_{i^*+1}(y)\chi_{\rm sr}(y_{i^*}^*,y\} \,
$$
and similarly for $i^*-1$. One can then extract the configuration recursively on the remaining variables, and this ensures that the final solution satisfy the ordering constraints.

\subsection{Viterbi decoding}
The two approaches above use the information containing on the single-site marginals $P_i$.
As a third approach, we propose to use Viterbi decoding to extract a solution from the output of the message-passing algorithm. 
This method has the advantage to take into account the information contained in the next-neighbors pairwise marginals $P_{i,i+1}$ and is therefore expected to be more efficient than the two previous method, which is confirmed by our results (see Fig.~\ref{fig:seqs_decodings}).
Viterbi decoding is a method to compute the configuration achieving the maximum of a probability distribution defined on a one-dimensional chain.
It would be an exact algorithm in the absence of long-range couplings: $J_{ij}=0$ for all $i,j$ with $|i-j|>1$.
Here, despite the presence of long-range coupling, we will use it as an heuristic.
The Viterbi algorithm builds the most-likely configuration 
\begin{align*}
    &\{y_1^*,\dots,y_L^*\}=\underset{{\bf y}}{\rm argmax}\{P^{\rm sr}({\bf y})\} \\
    &\text{with} \quad P^{\rm sr}({\bf y}) = P(y_1,y_2)\prod_{i=3}^L P(y_i|y_{i-1})
\end{align*}
in a recursive way from $i=1$ to $i=L$.
We introduce the following notations.
Let $p(y_i,i)$ be the probability of the most-likely configuration so far:
$$
    p(y_i,i) = P(y_1^*,y_2^*,\dots,y_{i-1}^*,y_i) \ ,
$$
and $s(y_i,i)=y_{i-1}^*$ be the last label found.
In the first step of the algorithm, we compute $p(y_2,2),s(y_2,2)$ for each value of $y_2\in\chi$:
\begin{align*}
    p(y_2,2) &= \underset{y_1}{\max}\{P_{12}(y_1,y_2)\} \\
    s(y_2,2) &= \underset{y_1}{\rm argmax}\{P_{12}(y_1,y_2)\}
\end{align*}
One then computes recursively the probabilities and labels $p(y_i,i),s(y_i,i)$, for each $i\in\{3, \dots, L\}$:
\begin{align*}
    p(x_i,i) &= \underset{y_{i-1}}{\max}\{p(y_{i-1},i-1)P(y_i|y_{i-1})\}\\
    s(y_i,i) &= \underset{y_{i-1}}{\rm argmax}\{p(y_{i-1},i-1)P(y_i|y_{i-1})\}\ .
\end{align*}
Finally, one obtains the configuration ${\bf y}^*$ recursively backward from $i=L$ to $i=1$:
\begin{align}
\begin{aligned}
    p &= \underset{y}{\max}\{p(y,L)\} = \underset{{\bf y}}{\max}\{P^{\rm sr}({\bf y})\} \\
    y_L^* &= \underset{y}{\rm argmax}\{p(y,L)\} \\
    y_i^* &= s(y_{i+1}^*,i+1) \quad \text{for each} \quad 1\leq i \leq L-1 \ .
\end{aligned}
\end{align}

\subsection{Viterbi sampling}
\begin{figure}[ht]
	\centering
	\includegraphics[width=1.0\columnwidth]{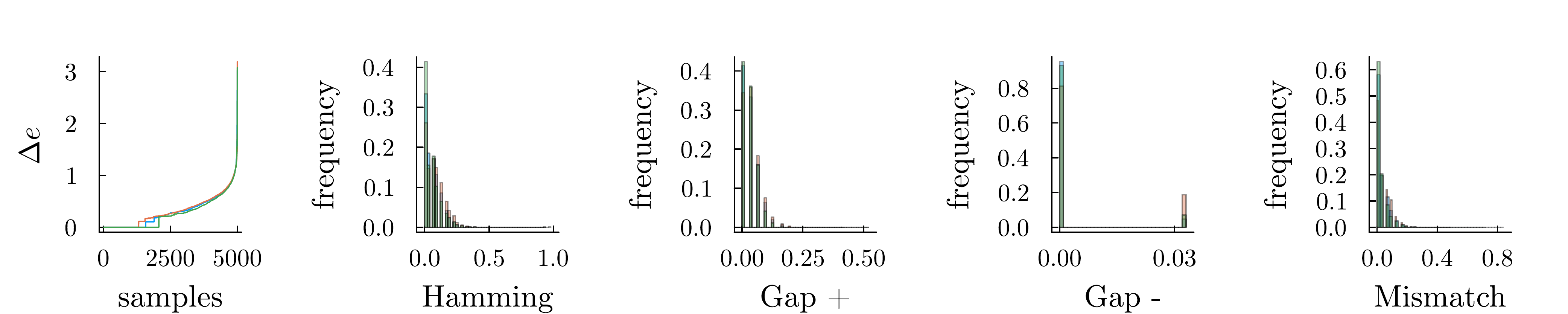} 
	\caption{{\bf Results of Viterbi Sampling}, obtained for the protein family PF00397.
		Each color/tone of gray corresponds to one sequence $\bfA$ extracted from the protein family.
		For each sequence, one computes the best alignment $\bfS^{\rm GS}$ (ground state) with our SCE algorithm (at $\beta=0.5$), and with Viterbi decoding.
		We sampled $5000$ sequences independently with our Viterbi sampling procedure, and compare them with $\bfS^{\rm GS}$.
		From left to right: 1. Difference in energy density between the sampled sequence $\bfS^{\rm samp}$ and the ground state $\Delta e = (\cH_{\rm DCA}(\bfS^{\rm samp}) - \cH_{\rm DCA}(\bfS^{\rm GS}))/L$, sorted by increasing values of $\Delta e$.
		2. Histogram of Hamming distances between $\bfS^{\rm samp}$ and $\bfS^{\rm GS}$.
		3. Histogram of Gap $+$, i.e. of the number of match states in $\bfS^{\rm GS}$ that have been replaced by a gap in $\bfS^{\rm samp}$.
		4. Histogram of Gap $-$, i.e. of the number of gap states in $\bfS^{\rm GS}$ that have been replaced by a match in $\bfS^{\rm samp}$.
		5. Histogram of Mismatches, i.e. of the number of times we have match states in both sequences, ground state and sampled, but corresponding to different amino acids.
	}
	\label{fig:viterbi_sampling}
\end{figure}

\begin{figure}[ht]
	\centering
	\includegraphics[width=1.0\columnwidth]{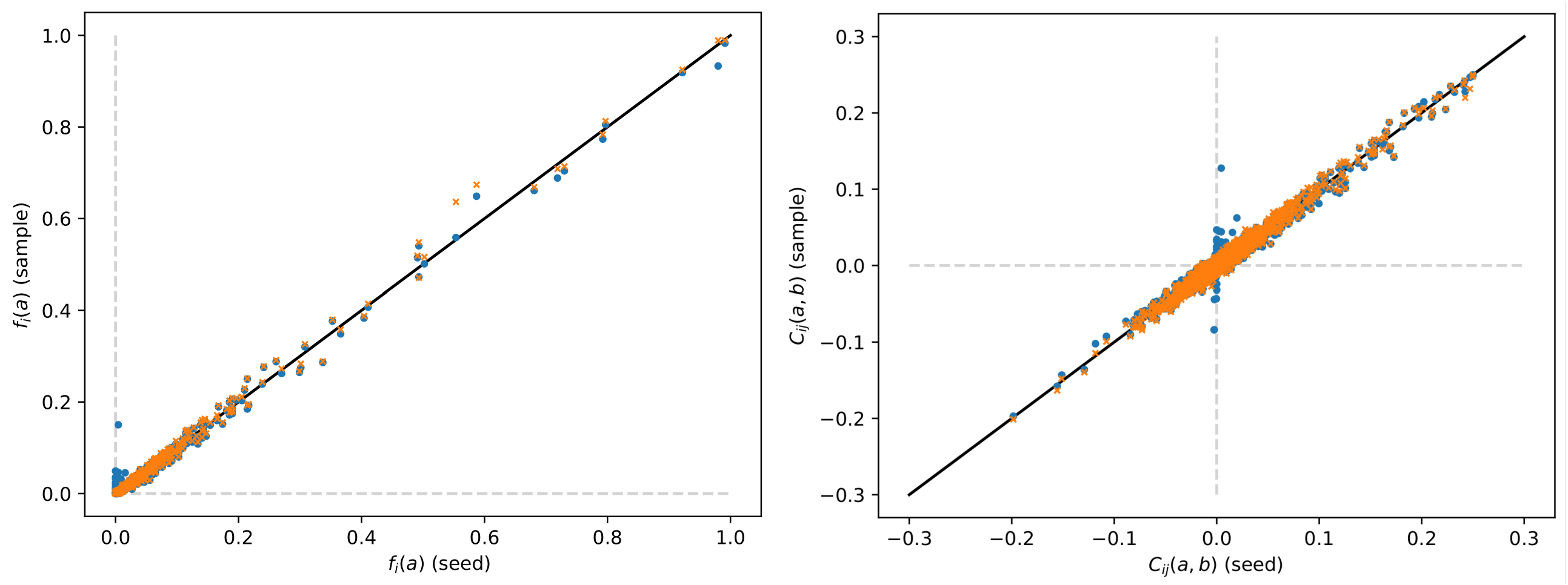}
	\caption{On family PF00397: Statistics of the sampled MSA $f_i^{\rm samp}, C_{ij}^{\rm samp}$ (blue/dark gray points) and of the MSA obtained by re-aligning each sequence of the seed with SCE+decoding $f_i^{\rm GS}, C_{ij}^{\rm GS}$ (orange/light gray crosses), plotted against the statistics of the seed MSA $f_i^{\rm seed}, C_{ij}^{\rm seed}$.
	}
	\label{fig:Vit-samp-freq-count}
\end{figure}

The Viterbi decoding strategy presented above can be turned into a sampling algorithm to generate configurations ${\bf y}$ sampled from the probability measure $P({\bf y})$ (see Eq.(\ref{eq:Boltzmann})), at a given inverse temperature $\beta$.
Configurations are produced recursively. 
One first picks the most polarized site:
$$i^*=\underset{i}{\rm argmax}\{\max_{y}(P_i(y)\}$$
and samples a value $y_{i^*}$ from its single-site marginal $P_{i^*}(y_{i^*})$.
Then nearest sites $i^*\pm 1$ are sampled from the conditional probabilities $P(y_{i^*\pm 1}|y_{i^*})$, and one proceeds in this way until the extremities of the chain.
Fig.~\ref{fig:viterbi_sampling}. shows the results of this sampling strategy on the protein family PF00397, for three sequences $\bfA$ randomly extracted from the protein family. 
One can see that the sampled sequences are close to the ground-state $\bfS^{\rm GS}$ (computed with Viterbi decoding), even coinciding with it for a fraction of them. 
Note that a few sampled sequences have a high $\Delta e$, i.e. these are not good alignments.
To further explore the quality of this sampling strategy, we have compared the first and second connected moments statistics of an MSA produced with Viterbi sampling against that obtained from the PFAM alignment. Results are shown in Fig.~\ref{fig:Vit-samp-freq-count}, on protein family PF00397. For each sequence $\bfS^{\rm seed}$ of the seed MSA, one considers the original sequence $\bfA$, and re-align it with our SCE+Viterbi decoding procedure, producing a new aligned sequence $\bfS^{\rm GS}$ (possibly coinciding with $\bfS^{\rm seed}$, see~\ref{subsec:comp_seed}). 
Then, one uses the SCE marginals to sample $\mathcal{N}=100$ sequences with the Viterbi sampling strategy described above.

We compute the statistics of the MSA made of the $M\mathcal{N}$ sampled sequences (with $M$ the number of sequences in the seed MSA). More precisely, we compute the one-site frequency count $f_i^{\rm samp}(a)$ (i.e. the frequency of observing amino acid $a$ at position $i$ in the MSA), and the correlations $C_{ij}^{\rm samp}(a,b)=f_{ij}^{\rm samp}(a,b)-f_i^{\rm samp}(a)f_j^{\rm samp}(b)$ (with $f_{ij}(a,b)$ the two-site frequency count).
We confront these statistics with the statistics $f_i^{\rm seed}, C_{ij}^{\rm seed}$ computed from the seed MSA (blue points), and see a good agreement between them, except for a small discrepancy observed at $f_i(a)\simeq 0$ and $C_{ij}(a,b)\simeq 0$.
For comparison, we also compute the statistics $f_i^{\rm GS}, C_{ij}^{\rm GS}$ of the MSA made with the re-aligned sequences $\bfS^{\rm GS}$ (orange crosses), and see that the statistics of the re-aligned MSA are comparable with the statistics of the sampled MSA, the latter being slightly more spread.

\section{Thermodynamic quantities}
\label{sec:thermo_quantities}
This small-coupling expansion has the advantage of providing an explicit expression for the free-entropy $\Phi=-\beta F$, expressed in terms of single site marginals $\{P_i\}_{i=1,\dots,L}$ and next-neighbors pairwise marginals $\{P_{e_i}\}_{i=1,\dots,L-1}$ (see equation~\ref{eq:free_energy_1storder}).
One can obtain an expression of the Bethe free-entropy $\Phi$ in terms of BP messages, by plugging the expression of the marginals (see equations~\ref{eq:BP_marginals}) in $\Phi^{\rm sr}$:
\begin{align*}
    \Phi^{\rm sr} &= \sum_{i=1}^{L-1}\log z_{e_i} -\sum_{i=1}^{L-1}\sum_{y,y'}P_{e_i}(y,y')\beta g_{i}(y,y') - \sum_{i=1}^{L-1}\sum_{y_i} P_i(y_i)\log F_i(y_i) \\
    &- \sum_{i=1}^{L-1}\sum_{y_{i+1}} P_{i+1}(y_{i+1})\log B_{i+1}(y_{i+1}) -\sum_{i=1}^L(1-d_i)\log z_i + \sum_{i=1}^L d_i \sum_{y}P_i(y)\beta H_i(y) \\
    &+ \sum_{i=2}^{L-1}\sum_y P_i(y)\beta f_i(y) + \sum_{i=2}^{L-1}\sum_y P_i(y)\log \widehat{F}_i(y_i) + \sum_{i=2}^{L-1}\sum_y P_i(y)\log \widehat{B}_i(y_i) \ ,
\end{align*}
where we have adopted the shorthand notation $e_i=(i,i+1)$. One can use the BP equations~(\ref{eq:BP_update}) to obtain the following identities:
\begin{align*}
    - \sum_{i=1}^{L-1}\sum_{y_i} P_i(y_i)\log F_i(y_i) = \sum_{i=1}^{L-1}\log z_{i\to e_i} - \sum_{i=1}^{L-1}\sum_{y_i}P_i(y_i)\beta H_i(y_i) - \sum_{i=2}^{L-1} \sum_{y_i}P_i(y_i)(\log \widehat{F}_i(y_i) + \beta f_i(y_i)) \ ,
\end{align*}
and:
\begin{align*}
    - \sum_{i=1}^{L-1}\sum_{y_{i+1}} P_{i+1}(y_{i+1})\log B_{i+1}(y_{i+1}) &= \sum_{i=1}^{L-1}\log z_{i+1\to e_i} -\sum_{i=1}^{L-1} \sum_{y_{i+1}}P_{i+1}(y_{i+1})\beta H_{i+1}(y_{i+1}) \\
    &-\sum_{i=1}^{L-2}\sum_{y_{i+1}}P_{i+1}(y_{i+1})\left(\log\widehat{B}_{i+1}(y_{i+1}) + \beta f_{i+1}(y_{i+1})\right) \quad .
\end{align*}
Using these identities one obtains the following expression for $\Phi^{\rm sr}$:
\begin{align*}
    \Phi^{\rm sr} &= \sum_{i=1}^{L-1}\log z_{e_i}  + \sum_{i=1}^L(1-d_i)\log z_i + \sum_{i=1}^{L-1}\log z_{i\to e_i} + \sum_{i=1}^{L-1}\log z_{i+1\to e_i} \\
    &- \beta\sum_{i=2}^{L-1}\sum_{y_i}P_i(y_i) f_i(y_i)-\beta \sum_{i=1}^{L-1}\sum_{y,y'}P_{e_i}(y,y')g_{i}(y,y')
\end{align*}
We can now use the following relations:
\begin{align*}
    z_i &= z_{i\to e_i}Z_{i,e_i}\\
    &=z_{i\to e_{i-1}}Z_{i,e_{i-1}} \ , \  \text{for} \quad i\in\{2, \dots, L-1\} \\
    z_1 &= z_{1\to e_1}Z_{1,e_1} \\
    z_L &= z_{L\to e_L}Z_{L,e_L}
\end{align*}
with:    
\begin{align*}
    Z_{i,e_i} &= \sum_{y_i}F_i(y_i)\widehat{B}_i(y_i) \\
    Z_{i,e_{i-1}} &= \sum_{y_i}B_i(y_i)\widehat{F}_i(y_i)
\end{align*}
to obtain a final expression for $\Phi^{\rm sr}$:
\begin{align*}
    \Phi^{\rm sr} &= \sum_{i=1}^{L-1} \log z_{e_i} + \sum_{i=1}^L\log z_i - \sum_{i=1}^{L-1}\log Z_{i,e_i}- \sum_{i=2}^L\log Z_{i,e_{i-1}}  \\
    &- \beta \sum_{i=2}^L\sum_{y_i}P_i(y_i)f_i(y_i)- \beta\sum_{i=1}^{L-1}\sum_{y_i, y_{i+1}}P_{e_i}(y_i,y_{i+1})g_{i}(y_i, y_{i+1})
\end{align*}
We finally recognize in the second line of this last equation the expression of $-\beta\mathcal{A}$. Indeed from the definition of the fields $f_i, g_{i}$ (see equation~(\ref{eq:def_external_fields})) one can check that 
\begin{align*}
    \mathcal{A} &= \sum_{i=2}^L\sum_{y_i}P_i(y_i)f_i(y_i) + \sum_{i=1}^{L-1}\sum_{y_i, y_{i+1}}P_{e_i}(y_i,y_{i+1})g_{i}(y_i, y_{i+1}) \ .
\end{align*}
The final expression for the free-entropy $\Phi=\Phi^{\rm sr}-\beta\mathcal{A}$ at first order in the small-coupling expansion is therefore just the usual expression for the Bethe free-entropy on the chain:
\begin{align}
\label{eq:Bethe_free_entropy}
    \Phi &= \sum_{i=1}^{L-1} \log z_{e_i} + \sum_{i=1}^L\log z_i - \sum_{i=1}^{L-1}\log Z_{i,e_i} - \sum_{i=2}^L\log Z_{i,e_{i-1}} , \
\end{align}
where:
\begin{align}
\begin{aligned}
    z_i &= \sum_{y_i}e^{\beta(H_i(y_i) +  f_i(y_i) ) }\widehat{F}_i(y_i)F_i(y_i) \ , \ \text{for} \quad i\in\{2,\dots, L-1\}\\
    z_1 &= \sum_{y_1} e^{\beta H_1(y_1)}\widehat{B}_1(y_1) \\
    z_L &= \sum_{y_L} e^{\beta H_L(y)}\widehat{F}_L(y_L)\\
    z_{e_i} &= \sum_{y_i, y_{i+1}}e^{\beta(J_{e_i}(y_i, y_{i+1}) +  g_{i}(y_i, y_{i+1}))} F_i(y_i)B_{i+1}(y_{i+1}) \ , \ \text{for} \quad i\in\{1,\dots, L-1\}\\
    Z_{i,e_i} &= \sum_{y_i}F_i(y_i)\widehat{B}_i(y_i) \ , \quad \text{for} \ i\in\{1,\dots,L-1\} \\
    Z_{i,e_{i-1}} &= \sum_{y_i}B_i(y_i)\widehat{F}_i(y_i)\ , \ \text{for} \quad i\in\{2,\dots, L\} \ .
\end{aligned}
\end{align}
The internal energy $U=\langle H({\bf y})\rangle$ can be expressed in terms of the BP marginals:
\begin{align}
\label{eq:internal_energy}
    U = -\sum_{i=1}^L\sum_{y_i}P_i(y_i)H_i(y_i) - \sum_{i < j}\sum_{y_i,y_j}p(y_i,y_j)J_{ij}(y_i,y_j)
\end{align}
where the joint-probability on any pair $i,j$ has been computed from from the set of single site marginals $P_i$ and nearest-neighbors $P_{e_i}$ recursively (see equations~\ref{eq:conditional_recursion}).
Finally, one can compute the entropy using the canonical identity:
\begin{align}
    S(\beta) = \Phi + U(\beta)/\beta \ ,
\end{align}
with $\Phi(\beta)$ and $U(\beta)$ computed respectively from (\ref{eq:Bethe_free_entropy}) and (\ref{eq:internal_energy}).

\bibliography{biblio.bib}

\begin{thebibliography}{10}

\bibitem{durbin1998}
R.~Durbin, S.~R. Eddy, A.~Krogh, and G.~Mitchison.
\newblock {\em Biological sequence analysis: probabilistic models of proteins
  and nucleic acids}.
\newblock Cambridge university press, 1998.

\bibitem{needleman1970}
S.~B. Needleman and C.~D. Wunsch.
\newblock A general method applicable to the search for similarities in the
  amino acid sequence of two proteins.
\newblock {\em Journal of molecular biology}, {\bfseries 48}(3), 443--453
  (1970).

\bibitem{smith1981}
T.~F. Smith and M.~S. Waterman.
\newblock Identification of common molecular subsequences.
\newblock {\em Journal of molecular biology}, {\bfseries 147}(1), 195--197
  (1981).

\bibitem{edgar2006}
R.~C. Edgar and S.~Batzoglou.
\newblock Multiple sequence alignment.
\newblock {\em Current Opinion in Structural Biology}, {\bfseries 16}(3),
  368--373 (2006).

\bibitem{altschul1997}
S.~F. Altschul, T.~L. Madden, A.~A. Sch{\"a}ffer, J.~Zhang, Z.~Zhang,
  W.~Miller, and D.~J. Lipman.
\newblock Gapped BLAST and PSI-BLAST: a new generation of protein database
  search programs.
\newblock {\em Nucleic acids research}, {\bfseries 25}(17), 3389--3402 (1997).

\bibitem{eddy2011}
S.~R. Eddy.
\newblock Accelerated profile HMM searches.
\newblock {\em PLoS computational biology}, {\bfseries 7}(10), e1002195 (2011).

\bibitem{el2019}
S.~El-Gebali, J.~Mistry, A.~Bateman, S.~R. Eddy, A.~Luciani, S.~C. Potter,
  M.~Qureshi, L.~J. Richardson, G.~A. Salazar, A.~Smart, et~al.
\newblock The Pfam protein families database in 2019.
\newblock {\em Nucleic acids research}, {\bfseries 47}(D1), D427--D432 (2019).

\bibitem{dejuan2013}
D.~De~Juan, F.~Pazos, and A.~Valencia.
\newblock Emerging methods in protein co-evolution.
\newblock {\em Nature Reviews Genetics}, {\bfseries 14}(4), 249--261 (2013).

\bibitem{cocco2018}
S.~Cocco, C.~Feinauer, M.~Figliuzzi, R.~Monasson, and M.~Weigt.
\newblock Inverse statistical physics of protein sequences: a key issues
  review.
\newblock {\em Reports on Progress in Physics}, {\bfseries 81}(3), 032601
  (2018).

\bibitem{marks2011}
D.~S. Marks, L.~J. Colwell, R.~Sheridan, T.~A. Hopf, A.~Pagnani, R.~Zecchina,
  and C.~Sander.
\newblock Protein 3D Structure Computed from Evolutionary Sequence Variation.
\newblock {\em PLOS ONE}, {\bfseries 6}(12), 1--20 (2011).

\bibitem{morkos2011}
F.~Morcos, A.~Pagnani, B.~Lunt, A.~Bertolino, D.~S. Marks, C.~Sander,
  R.~Zecchina, J.~N. Onuchic, T.~Hwa, and M.~Weigt.
\newblock Direct-coupling analysis of residue coevolution captures native
  contacts across many protein families.
\newblock {\em Proceedings of the National Academy of Sciences}, {\bfseries
  108}(49), E1293--E1301 (2011).

\bibitem{procaccini2011}
A.~Procaccini, B.~Lunt, H.~Szurmant, T.~Hwa, and M.~Weigt.
\newblock Dissecting the Specificity of Protein-Protein Interaction in
  Bacterial Two-Component Signaling: Orphans and Crosstalks.
\newblock {\em PLOS ONE}, {\bfseries 6}(5), 1--9 (2011).

\bibitem{baldassi2014}
C.~Baldassi, M.~Zamparo, C.~Feinauer, A.~Procaccini, R.~Zecchina, M.~Weigt, and
  A.~Pagnani.
\newblock Fast and Accurate Multivariate Gaussian Modeling of Protein Families:
  Predicting Residue Contacts and Protein-Interaction Partners.
\newblock {\em PLOS ONE}, {\bfseries 9}(3), 1--12 (2014).

\bibitem{feinauer2016}
C.~Feinauer, H.~Szurmant, M.~Weigt, and A.~Pagnani.
\newblock Inter-Protein Sequence Co-Evolution Predicts Known Physical
  Interactions in Bacterial Ribosomes and the Trp Operon.
\newblock {\em PLOS ONE}, {\bfseries 11}(2), 1--18 (2016).

\bibitem{figliuzzi2015}
M.~Figliuzzi, H.~Jacquier, A.~Schug, O.~Tenaillon, and M.~Weigt.
\newblock {Coevolutionary Landscape Inference and the Context-Dependence of
  Mutations in Beta-Lactamase TEM-1}.
\newblock {\em Molecular Biology and Evolution}, {\bfseries 33}(1), 268--280
  (2015).

\bibitem{morcos2016}
R.~R. Cheng, O.~Nordesjö, R.~L. Hayes, H.~Levine, S.~C. Flores, J.~N. Onuchic,
  and F.~Morcos.
\newblock {Connecting the Sequence-Space of Bacterial Signaling Proteins to
  Phenotypes Using Coevolutionary Landscapes}.
\newblock {\em Molecular Biology and Evolution}, {\bfseries 33}(12), 3054--3064
  (2016).

\bibitem{hopf2017}
T.~A. Hopf, J.~B. Ingraham, F.~J. Poelwijk, C.~P. Sch{\"a}rfe, M.~Springer,
  C.~Sander, and D.~S. Marks.
\newblock Mutation effects predicted from sequence co-variation.
\newblock {\em Nature biotechnology}, {\bfseries 35}(2), 128--135 (2017).

\bibitem{trinquier2021}
J.~Trinquier, G.~Uguzzoni, A.~Pagnani, F.~Zamponi, and M.~Weigt.
\newblock Efficient generative modeling of protein sequences using simple
  autoregressive models.
\newblock {\em Nature communications}, {\bfseries 12}(1), 1--11 (2021).

\bibitem{Plefka82}
T.~Plefka.
\newblock Convergence condition of the {TAP} equation for the infinite-ranged
  Ising spin glass model.
\newblock {\em Journal of Physics A: Mathematical and General}, {\bfseries
  15}(6), 1971--1978 (1982).

\bibitem{eddy2020}
G.~W. Wilburn and S.~R. Eddy.
\newblock Remote homology search with hidden Potts models.
\newblock {\em PLOS Computational Biology}, {\bfseries 16}(11), 1--22 (2020).

\bibitem{Muntoni2020}
A.~P. Muntoni, A.~Pagnani, M.~Weigt, and F.~Zamponi.
\newblock Aligning biological sequences by exploiting residue conservation and
  coevolution.
\newblock {\em Phys. Rev. E}, {\bfseries 102}, 062409 (2020).

\bibitem{talibart2021}
H.~Talibart and F.~Coste.
\newblock PPalign: optimal alignment of Potts models representing proteins with
  direct coupling information.
\newblock {\em BMC bioinformatics}, {\bfseries 22}(1), 1--22 (2021).

\bibitem{ovchinnikov2021}
S.~Petti, N.~Bhattacharya, R.~Rao, J.~Dauparas, N.~Thomas, J.~Zhou, A.~M. Rush,
  P.~Koo, and S.~Ovchinnikov.
\newblock {End-to-end learning of multiple sequence alignments with
  differentiable Smith–Waterman}.
\newblock {\em Bioinformatics}, {\bfseries 39}(1) (2022).
\newblock btac724.

\bibitem{YeGe91}
A.~Georges and J.~S. Yedidia.
\newblock How to expand around mean-field theory using high-temperature
  expansions.
\newblock {\em Journal of Physics A: Mathematical and General}, {\bfseries
  24}(9), 2173--2192 (1991).

\bibitem{OpMa01}
M.~Opper and D.~Saad.
\newblock {\em From Naive Mean Field Theory to the TAP Equations}.
\newblock MIT Press, 2001.

\bibitem{PagParRat03}
A.~Pagnani, G.~Parisi, and M.~Rati\'eville.
\newblock Near-optimal configurations in mean-field disordered systems.
\newblock {\em Phys. Rev. E}, {\bfseries 68}, 046706 (2003).

\bibitem{MaPaRi02}
E.~Marinari, A.~Pagnani, and F.~Ricci-Tersenghi.
\newblock Zero-temperature properties of RNA secondary structures.
\newblock {\em Phys. Rev. E}, {\bfseries 65}, 041919 (2002).

\bibitem{Rfam_17}
I.~Kalvari, J.~Argasinska, N.~Quinones-Olvera, E.~P. Nawrocki, E.~Rivas, S.~R.
  Eddy, A.~Bateman, R.~D. Finn, and A.~I. Petrov.
\newblock {Rfam 13.0: shifting to a genome-centric resource for non-coding RNA
  families}.
\newblock {\em Nucleic Acids Research}, {\bfseries 46}(D1), D335--D342 (2017).

\bibitem{Eddy_homologyfailure}
C.~M. Weisman, A.~W. Murray, and S.~R. Eddy.
\newblock Many, but not all, lineage-specific genes can be explained by
  homology detection failure.
\newblock {\em PLOS Biology}, {\bfseries 18}(11), 1--24 (2020).

\bibitem{MeMo09}
M.~M\'ezard and A.~Montanari.
\newblock {\em Physics, Information, Computation}.
\newblock Oxford University Press, 2009.

\end{thebibliography}

\end{document}